\documentclass{aa}
\usepackage{times}
\usepackage{astron}
\usepackage{psfig}
\usepackage{changebar}
\begin{document}
\bibliographystyle{astron}
\newcommand{\adsn}{\textit{Abell\,370}}
\newcommand{\evvs}{\textit{Cl\,1447$+$26}}
\newcommand{\ndnd}{\textit{Cl\,0303$+$17}}
\newcommand{\nndn}{\textit{Cl\,0939$+$47}}
\newcommand{\nnes}{\textit{Cl\,0016$+$16}}
\newcommand{\ea}{E$+$A}
\newcommand{\hst}{\textit{Hubble Space Telescope}}
\newcommand{\kpc}{$R_e$ in $\mathrm{kpc}$}
\newcommand{\magn}{\ensuremath{\mathrm{mag}}}
\newcommand{\me}{\ensuremath{^{-1}}}
\hyphenation{Kor-men-dy}

\nochangebars

   \thesaurus{3(11.03.1; 11.05.1; 11.05.2; 11.06.1; 11.06.2)} 
   \title{Probing early-type galaxy evolution with the Kormendy 
   relation\thanks{Based on observations with the NASA/ESA
               {\it Hubble Space Telescope}, obtained at the Space Telescope
               Science Institute, which is operated by AURA, Inc., under
               NASA contract NAS 5-26555.} }

   \author{B. L. Ziegler \inst{1} \thanks{{\it Present Address:}
     Department of Physics, South Road, Durham DH1 3LE, United Kingdom}      
       \and R. P. Saglia\inst{1}, R. Bender\inst{1},
     P. Belloni\inst{1}\thanks{Visiting astronomer of the 
               German--Spanish Astronomical Center, Calar Alto, operated by 
               the Max--Planck--Institut f\"ur Astronomie, Heidelberg, jointly
               with the Spanish National Commission for Astronomy.} 
       \and L. Greggio\inst{1}\inst{2} \and S. Seitz\inst{1} }

   \offprints{B. L. Ziegler}

   \institute{{Universit\"ats--Sternwarte, 
   Scheinerstra\ss e 1, 81679~M\"unchen, Germany\\
              email: ziegler@usm.uni-muenchen.de}
   \and{Dipartimento di Astronomia, via Zamboni 33, I40100 Bologna, Italy}}

   \date{Received June 1998 / Revised Jan. 1999 / Accepted 1999}

   \maketitle

   \begin{abstract}

     We investigate the evolution of early-type galaxies in four
     clusters at $z=0.4$ (\adsn, \ndnd, \nndn\ and
     \textit{Cl\,1447$+$26}) and in one at $z=0.55$ (\nnes). The
     galaxies are selected according to their spectrophotometrically
     determined spectral types and comprise the morphological classes
     E, S0 and Sa galaxies.  Structural parameters are determined by a
     two-component fitting of the surface brightness profiles derived
     from HST images.  Exploring a realistic range of K-corrections
     using Bruzual and Charlot models, we construct the rest-frame
     $B$-band Kormendy relations
     ($\mathrm{\langle\-\mu_e}\rangle-\log(R_e)$) for the different
     clusters.  \cbstart We do not detect a systematic change of the
     slope of the relation as a function of redshift. We discuss in
     detail how the luminosity evolution, derived by comparing the
     Kormendy relations of the distant clusters with the local one for
     Coma, depends on various assumptions and give a full description
     of random and systematic errors by exploring the influences of
     selection bias, different star formation histories and
     K-corrections.  \cbend{}
     
     Early-type galaxies with modest disk components (S0 and Sa) 
     do not differ significantly in their evolution from disk-less
     ellipticals. 

     The observed luminosity evolution is compatible with pure passive
     evolution models (with redshift of formation $z>2$) but also with
     models that allow ongoing star formation on a low level, like 
     exponentially decaying star formation models with an e-folding 
     time of $\tau=1$ Gyr. 


   \end{abstract}


\section{Introduction}
\label{intro}

In the past years, many observations have been made to investigate the
redshift evolution of elliptical galaxies and to compare them with
stellar population synthesis models. Most of the authors conclude that
the stellar populations in cluster ellipticals evolve mainly in a
passive manner \cite[and
others]{BLE92,AECC93,RS95,BAECSS96,BZB96,ESDCOBS97,SED98,ZB97}. Passive
evolution models assume a short but intensive initial star formation
phase and no subsequent star formation \cite{BC93}. Other
studies have shown that most of the observations are also compatible
with hierarchical evolution models \cite{Kauff96,KC98}. One of the most
accurate tools to test galaxy evolution is offered by the scaling
relations which hold for elliptical galaxies, like the
Fundamental Plane \cite{DD87,DLBDFTW87}. Here we write the Fundamental
Plane equation in a form where the mean effective surface brightness
$\mathrm{\langle \mu_e\rangle}$ is given as a function
of effective radius $R_e$ (in kpc) and velocity dispersion $\sigma$:
\begin{equation}
\label{FP}
\mathrm{\langle \mu_e\rangle} = \tilde{a} + \tilde{b} \cdot \log R_e + 
c \cdot \log \sigma
\end{equation}
First observations of the Fundamental Plane at intermediate redshifts
indicate indeed the passive evolution of elliptical cluster galaxies
\cite{DF96,KDFIF97,JH97,BSZBBGH97b,DFKI98}. 

The determination of the Fundamental Plane parameters at even modest
redshifts is non-trivial and requires good signal-to-noise ratios.
The velocity dispersion can only be derived from
intermediate-resolution spectra obtained with either 8m-class telescopes or
very long exposure times at 4m class telescopes \cite{ZB97,KDFIF97}. Because
the galaxy size is of order of a few arcsec at $z>0.2$, the structural
parameters can be measured accurately only in the spatially
highly resolved \hst\ images. With WFPC2 delivering such images now in
great numbers, but lacking the spectroscopic information, many studies
have been made exploiting the projection of the Fundamental Plane onto
the plane defined by $R_e$ and $\mathrm{\langle \mu_e\rangle}$, i.e. 
the Kormendy relation \cite{Korme77}:
\begin{equation}
\label{KR}
\mathrm{\langle \mu_e\rangle} = a + b \cdot \log R_e
\end{equation}
This relation was used to perform the Tolman test for the cosmological
dependence of the surface brightness assuming passive luminosity
evolution for elliptical galaxies \cite{PDC96,MCKFB98}. While the first
group finds the $(1+z)^4$ dependence of the surface brightness in an
expanding Universe confirmed, the second group points out that the
scatter in the observed data is too large to significantly constrain
any cosmological model. Other groups utilized the Kormendy relation to
investigate the luminosity evolution itself both for field galaxies
\cite{SCYLE96,FCAF98} and for cluster galaxies
\cite{BSL96,SBL97,BASECDOPS98}. All these studies conclude that the
evolution of the stellar populations in spheroidal galaxies is most
probably purely passive at low redshifts ($z\la0.6$) and that their
formation epoch lies at high redshift ($z_f>2$).

Most of the cited studies have however the disadvantage that they must
rely on photometry only, so that neither cluster membership of a
galaxy is guaranteed, nor that the sample is not contaminated by some
post-starburst galaxies like \ea\ galaxies. The early-type galaxies
are also not distinguished with respect to E or S0 types. All the
authors assume a fixed slope $b$ of the Kormendy relation, although
its validity at any redshift is not proven a priori. The errors in the
transformation from HST magnitudes into the photometric system of the
local reference system are not always taken into account in the
derivation of the luminosity evolution. All these points are addressed
in this paper.  We start our investigations of spectrophotometrically
defined early-type member galaxies in five distant clusters
(Sect.\,\ref{sample}) with a thorough analysis of the possible
systematic errors arising from the magnitude calibration
(Sect.\,\ref{seccal}). After examining the coefficients of the
Kormendy relation of some representative local samples, we determine
its slope in the distant clusters by a free bisector fit and derive
the luminosity evolution by comparison with one specific local cluster
sample. Then, we fit all the cluster samples with the same slope for
the Kormendy relation and study the difference in the derived
evolution (Sect.\,\ref{lumev}). The influence of a number of
parameters is investigated in Sect.\,\ref{para}. We also look at the
results for subsamples containing only galaxies with and without a
substantial disk component, and for the whole sample augmented by a
few known \ea\ galaxies. Finally, we investigate which evolutionary
models (not only the passive one) can fit the data within their errors
(Sect.\,\ref{model}).  \cbstart In the appendix, we present the
photometric parameters of all the galaxies in the distant clusters
studied here.  \cbend{}


\section{The sample and parameter determination}
\label{sample}

In this paper we examine the early-type galaxy population in four
clusters at redshifts around $z=0.4$ (\adsn, \ndnd, \nndn\ and
\textit{Cl\,1447$+$26}) and one at $z=0.55$ (\nnes). From ground-based
spectrophotometry we determined cluster membership and spectral type of
the galaxies, whereas HST images were used to derive morphological and
structural parameters.

With the exception of \adsn, all clusters were observed at the 3.5m
telescope of the Calar Alto Observatory. Images were taken in the
broad-band filters $B$, $R$ and $I$ and in eight different narrow-band
filters, which were chosen to sample characteristic features of galaxy
spectra taking into account the clusters' redshifts. From the
multi-band imaging, low resolution spectral energy distributions were
constructed which were fitted by template spectra of local galaxy types
\cite{CWW80}.  Special care was taken to find post-starburst (\ea)
galaxies. Their existence was revealed by a good fit of their SED by
one of six different model spectra synthesized by the superposition of
an elliptical and a burst component. In this manner, cluster membership
could be determined with good accuracy and galaxies were classified as
either early-type (ET), spiral (Sbc or Scd), irregular (Im) or
post-starburst (\ea) \cite[\cbstart where numerous SED fitting
examples can be found\cbend{}]{BBTR95,BR96,Vulet96,BVR97}.  Thus, galaxies
were selected according to spectral type and, in the following study,
only cluster members of type ET were included.  Morphologically, these
galaxies could be either E, S0 or Sa galaxies. In the case of \adsn, we
include only spectroscopically confirmed ET member galaxies
\cite{MSFM88,PK91,ZB97}.

\cbstart HST-WFPC2 $R$ images do exist of the cores of the clusters
\adsn, \evvs, \nndn\ and \ndnd, whereas \nnes\ was observed both in $V$
and $I$ (see Table\,\ref{gals}). Additionally, an outer region of
\adsn\ and of \nndn\ was observed in $V$ and $I$, too. \cbend{} As
expected from the density-morphology relation \cite{Dress80,DOCSE97},
only a small number of ET galaxies are found in the outer fields,
whereas the core images contain 30 to 40 ET galaxies of our
ground-based sample. \cbstart Due to the uncertainties affecting the
photometric calibration (see Sect.\,\ref{seccal}), we did not combine
the $V$ and $I$ data of the same galaxies transformed to $B_{\mathrm
rest}$, \cbend{} and we exclude from our statistical investigation those
samples which have less than 10 galaxies.


With the exception of \adsn, the WFPC2 images were retrieved from the
ST/ECF archive as re-processed frames using up to date reduction
files.  In the case of \adsn, our original HST data of the core of the
cluster were used. The individual images per filter were combined
using the \textsf{imshift} and \textsf{crrej} tasks within the
\textsf{IRAF stsdas} package \cite{HST_DHB}. The candidate galaxies
were then extracted, stars and artifacts removed, a sky value assigned
and the surface-brightness profile fitted \cite{Flech97} within MIDAS
\cite{MIDAS}. The profile analysis followed the prescription described
by Saglia et al.  \cite*{SBBBCDMW97}. In short, a PSF (computed using
the Tinytim program) broadened $r^{1/4}$ and an exponential component
were fitted simultaneously and separately to the circularly averaged
surface brightness profiles. The quality of the fits were explored by
Monte Carlo simulations, taking into account sky-subtraction
corrections, the signal-to-noise ratio, the radial extent of the
profiles and the $\chi^2$ quality of the fit. \cbstart In this way, we
were able to detect the disk of lenticular S0 and Sa galaxies and
larger disky ellipticals and to derive not only the \textit{global}
values of the total magnitude $M_{\mathrm{tot}}$ and the effective
radius $r_e$ (in arcsec), but also the luminosity and scale of the
bulge ($m_b$ and $r_{e,b}$) and disk ($m_d$ and $h$) component
separately, within the limitations described by Saglia et al.
\cite*[especially Fig.\,13]{SBBBCDMW97}. Extensive tests have been
made in that paper and it was shown that the fits have
only problems with nearly edge-on galaxies. Since all the investigated
galaxies have low ellipticities, the deviations around $\log R_e - 0.3
\langle \mu_e\rangle$, which is nearly parallel to the Kormendy
relation, are minimal. The average error in $M_{\mathrm{tot}}$ is
$0.15\,\magn$ and $25\%$ in $r_e$. All the photometric parameters of
each galaxy are given in the Tables of the Appendix, although only the
global values were used to construct the Kormendy relations. \cbend{}
Galaxies with $r_e<0.25\,\arcsec$ were rejected from our sample
because in this case only 5 or less pixels would contribute to the
bulge. The number of early-type galaxies (ET) of the different
clusters in the observed filters remaining for our investigations are
listed in Table\,\ref{gals}.

The samples are therefore characterized by a conservative selection,
because we pick up all E, S0 and Sa galaxies. However, having derived
the disk-to-bulge ratios for the clusters' galaxies, in a second step
we analyze subsamples of objects with $d/b\leq0.2$ (called E in the
following) and $d/b>0.2$ (called S0, but could include also Sa).

\begin{table*}[htb]
\caption[]{The sample. Col.\,1: cluster name used here, Col.\,2: HST
filter, Cols.\,3 -- 5: number of galaxies, Col.\,6: $B_{\mathrm{tot}}$
magnitude cut-off, Col.\,7: absolute $B$ magnitude limit 
($H_0=60\,\mathrm{km\,s\me\,Mpc\me}, q_0=0.1$), Cols.\,8 -- 10: 
minimum, median and maximum of the $\log R_e$ distribution.}
\label{gals}
\begin{flushleft}
\begin{tabular}{llrrcccccc}
\noalign{\smallskip}
\hline
\noalign{\smallskip}
 cluster & filter & E,S0,Sa & S0,Sa & \ea & $B_{\mathrm{lim}}$ & 
$M_{B,\mathrm{lim}}$ & Min($\log R_e$) & Med($\log R_e$) & Max($\log R_e$) \\
         &        &         &       &     & \magn              & 
\magn & \kpc            & \kpc          & \kpc          \\ 
\noalign{\smallskip}
\hline
\noalign{\smallskip}
ComaSBD & B     & 39 & 14 & 0 & 16.55 & $-$18.77 & 0.04 & 0.41 & 1.55 \\
  a370v & F555W &  9 &  1 & 0 & 20.87 & $-$20.82 & 0.44 & 0.63 & 0.92 \\
  a370r & F675W & 17 &  8 & 2 & 21.25 & $-$20.44 & 0.33 & 0.67 & 1.71 \\
  a370i & F814W &  9 &  3 & 0 & 21.32 & $-$20.37 & 0.43 & 0.56 & 0.89 \\
cl1447r & F702W & 31 & 11 & 2 & 22.35 & $-$19.43 & 0.21 & 0.42 & 1.07 \\
cl0939v & F555W &  8 &  2 & 3 & 22.19 & $-$19.70 & 0.28 & 0.52 & 0.58 \\
cl0939r & F702W & 26 & 16 & 9 & 23.22 & $-$18.67 & 0.19 & 0.42 & 1.13 \\
cl0939i & F814W &  6 &  2 & 2 & 22.33 & $-$19.56 & 0.25 & 0.48 & 0.61 \\
cl0303r & F702W & 24 & 10 & 6 & 22.20 & $-$19.75 & 0.19 & 0.53 & 1.13 \\
cl0016v & F555W & 30 &  7 & 7 & 22.13 & $-$20.52 & 0.27 & 0.49 & 1.40 \\
cl0016i & F814W & 28 &  9 & 3 & 22.54 & $-$20.11 & 0.27 & 0.51 & 1.38 \\
\noalign{\smallskip}
\hline
\end{tabular}
\end{flushleft}
\end{table*}


\section{Calibration of the data}
\label{seccal}

\begin{figure}[htb]
\psfig{figure=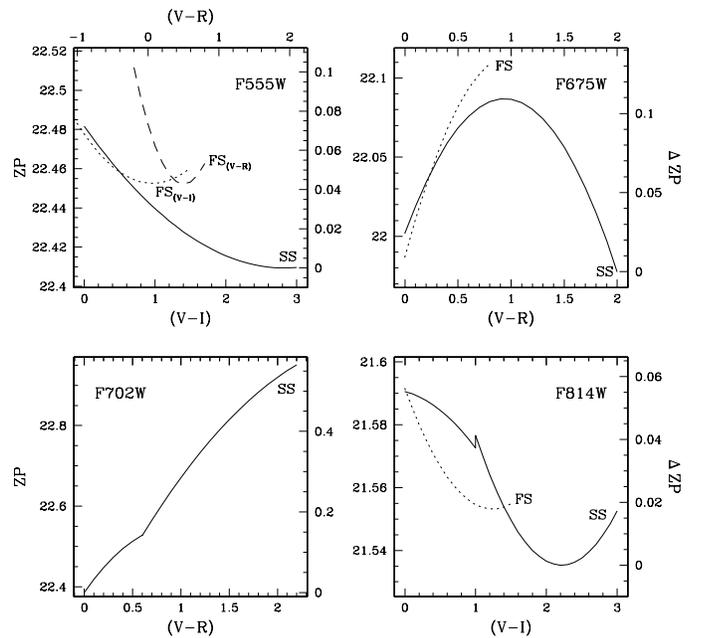,width=8.8cm}
\caption[]{Transformation between HST-WFPC2 filter magnitudes and
  Johnson-Cousins filter magnitudes. Zeropoints (ZP) on the left y-axis
  are calculated according to Table\,10 of Holtzman et al. (including
  the gain ratio of $0.753$\,\magn), $\Delta $ZP on the right y-axis is
  the offset from the minimum ZP. Dashed lines correspond to the
  `flight system' (FS), solid lines to the `synthetic system' (SS). In
  the case of the F555W filter, FS zeropoints are given as a function
  of $(V-I)$ color (lower x-axis) and $(V-R)$ color (upper x-axis),
  respectively, and x-axes zeropoints are shifted in order to match the
  same typical galaxy colors on a vertical line.  }
\label{trafo}
\end{figure}

\cbstart The mean effective surface brightness $\mathrm{\langle\mu_e\rangle_i}$
within the (global) effective radius $r_e$ for a given
HST filter ($i=V,R,I$) is defined as (cf. with Eq. (9) of Holtzman et
al.  \cite*{HBCHTWW95}):
\begin{equation}
\mathrm{\langle \mu_e\rangle _i}  = 
- 2.5 \log \left( I(<r_e) / \pi r_e^2 \right) + \mathrm{ZP_i(color)},
\end{equation}
where $I(<r_e)$ is the measured flux inside an aperture of radius
$r_e$:
\begin{eqnarray}
\label{sbhst}
\langle\mu_e\rangle_i = & - & 2.5 \log(\mathrm{counts}) + 2.5 \log t
+ 2.5 \log\mathrm{GR} \nonumber\\
& + & 2.5 \log (\pi) + 5 \log r_e + \mathrm{ZP_i(color)} 
\end{eqnarray}
%
We use a gain ratio of GR~$=2.0$ for all WFPC2 chips, since the
differences between the chips result in magnitude differences which are
much smaller than the other systematic errors. \cbend{} To be able to
compare the HST data of the distant galaxies with local samples we
transform the HST magnitudes into the standard Johnson-Cousins
photometric system, in which most ground-based studies are
accomplished. For the transformation between WFPC2 magnitudes and
UBVRI, there exist two systems, the `flight system' and the `synthetic
system' \cite{HBCHTWW95}. The `flight system' is based on measurements
of standard stars with colors $(V-I)<1.5$, whereas the `synthetic
system' is calculated from an atlas of stellar spectra.  Since
elliptical galaxies have colors redder than the stars used for the
calibration of the 'flight system', we rely on the `synthetic system'.
It is worth noticing that there are systematic differences between the
two systems in the overlapping color range. As shown in
Fig.\,\ref{trafo} these differences are small: $\Delta m\leq0.005,
0.02, 0.01$\,\magn\ for the F814W, F675W, F555W filter, respectively.
In Table\,\ref{calib}, we quote the zeropoints ZP$_1$ for an exposure
time of 1s for colors $(V-R)=1.25$ (F675W and F702W) and $(V-I)=2.2$
(F555W and F814W). The errors refer to variations in colors within the
ranges $1.0\leq(V-R)\leq1.5$ and $1.8\leq(V-I)\leq2.6$, which are
appropriate for early type galaxies. \cbstart The
zeropoints ZP$_t$ for the relevant exposure times $t$ (GR is already
included in both ZP) as well as the total integration time $t_{\mathrm
  tot}$ are also given. We notice that $t$ is the integration time of
both a single exposure and of the median image as constructed by the
task \textsf{crrej} (old version), whereas $t_{\mathrm tot}$ is the sum of all
individual exposures. The ZP are calculated using the coefficients of
Table\,10 of Holtzman et al.  \cite*{HBCHTWW95}. We checked the ZP of
\adsn\ by comparing the HST growth curves of eight galaxies to those
obtained with NTT data \cite{Ziegl98a} and find agreement within 
$\approx 0.01\,\magn$. The ZPs of the Calar Alto data of the other clusters are
not determined to better than $\approx 0.02\,\magn$. \cbend{}

\begin{table*}[hbt]
\caption[]{Calibration of $\mathrm{\langle \mu_e\rangle}$.  Col.\,1:
cluster name used here, Col.\,2: HST filter, Col.\,3: redshift, 
Col.\,4: zeropoint for 1s exposure, the error is the maximum deviation
of ZP in the assumed color range of the galaxies, Col.\,5: exposure time of
individual HST frames, Col.\,6: ZP for this exposure time, Col.\,7: total
integration time, Cols.\,8. \& 9.: reddening, Cols.\,10. \& 11.:
extinction, Col.\,12: mean K-correction and
maximum deviations for our model SEDs. }
\label{calib}
\begin{flushleft}
\begin{tabular}{llllllrllllr}
\noalign{\smallskip}
\hline
\noalign{\smallskip}
 cluster & filter & z & ZP$_1$ & t & ZP$_t$ & $t_{\mathrm tot}$ & 
$E_{(B-V)}$ & $E_{(B-V)}$ & $A_i$ & $A_i$ & $K(B,i)$\\
         &        &   &        &   &        &                   &
BH          & SFD         & RL    & SFD   &         \\
\noalign{\smallskip}
\hline
\noalign{\smallskip}
Coma    &   B   & 0.024 &                &      &       &       &
0.0103 & 0.0089 & 0.042 & 0.038 & $0.12\pm0.02$\\
  a370v & F555W & 0.375 & $22.41\pm0.01$ & 1000 & 29.91 &  8000 &
0.0122 & 0.0384 & 0.038 & 0.127 & $-0.25\pm0.05$\\
  a370r & F675W & 0.375 & $22.08\pm0.02$ & $^{\dag}$ &  &  5600 &
0.0122 & 0.0384 & 0.028 & 0.103 & $1.02\pm0.04$\\
  a370i & F814W & 0.375 & $21.54\pm0.00$ & 2100 & 29.84 & 12600 &
0.0122 & 0.0384 & 0.018 & 0.074 & $1.88\pm0.08$\\
cl1447r & F702W & 0.389 & $22.74\pm0.07$ & 2200 & 31.11 &  4200 &
0.0202 & 0.0340 & 0.047 & 0.091 & $1.00\pm0.02$\\
cl0939v & F555W & 0.407 & $22.41\pm0.01$ & 1000 & 29.91 &  8000 &
0.0042 & 0.0164 & 0.013 & 0.054 & $-0.39\pm0.06$\\
cl0939r & F702W & 0.407 & $22.74\pm0.07$ & 2100 & 31.05 & 21000 &
0.0042 & 0.0164 & 0.010 & 0.044 & $0.95\pm0.03$\\
cl0939i & F814W & 0.407 & $21.54\pm0.00$ & 2100 & 29.84 & 10500 &
0.0042 & 0.0164 & 0.006 & 0.032 & $1.84\pm0.06$\\
cl0303r & F702W & 0.416 & $22.74\pm0.07$ & 2100 & 31.05 & 12600 &
0.0892 & 0.1326 & 0.206 & 0.354 & $0.93\pm0.02$\\
cl0016v & F555W &  0.55 & $22.41\pm0.01$ & 2100 & 30.72 & 12600 &
0.0232 & 0.0572 & 0.072 & 0.190 & $-0.93\pm0.09$\\
cl0016i & F814W &  0.55 & $21.54\pm0.00$ & 2100 & 29.84 & 16800 &
0.0232 & 0.0572 & 0.035 & 0.111 & $1.62\pm0.05$\\
\noalign{\smallskip}
\hline
\end{tabular}

$^{\dag}$ The exposure times of the individual frames of \textit{a370r} 
have not a single value.
\end{flushleft}
\end{table*}

In order to be able to compare the data of the clusters, which are at
different redshifts and have been observed in different filters, all
observed magnitudes are converted to restframe $B$ magnitudes and
corrected for the cosmological dimming of the surface brightness. In
addition to applying the K-corrections ($K$), the galactic extinction
($A$) in the respective band ($i=V,R$ or $I$) has to be subtracted:
\begin{equation}
\label{sbrest}
\mathrm{\langle \mu_e\rangle}_B  = \mathrm{\langle \mu_e\rangle}_i 
- A_i + K(B,i,z) 
\end{equation}
\begin{equation}
\label{sbcor}
\mathrm{\langle \mu_e\rangle_{cor}} = \mathrm{\langle \mu_e\rangle}_B
- 10 \log(1+z) 
\end{equation}

There are two major sources for reddening values of the Galaxy in the
literature: one is based on H\,\textsc{i} measurements \cite[BH]{BH84},
the other on COBE/DIRBE and IRAS/ISSA FIR data \cite[SFD]{SFD98}.
Because there are systematic differences, in Table\,\ref{calib} we list
$E(B-V)$ values derived from both methods. For the conversion from
$E(B-V)$ into $A_i$ there exist different interstellar extinction laws
in the literature.  In Table\,\ref{calib}, the extinction is calculated
once using the law by Rieke and Lebofsky \cite*[RL]{RL85} (according to
their Table\,3) and BH reddenings, and a second time using SFD
reddenings according to their extinction curve (their Table\,6, Landolt
filters).  The absorption coefficients $A_i$ derived from SFD are
systematically larger by about $0.06$\,\magn\ (up to $0.15$\,\magn) with
respect to those derived with the other prescriptions.

\begin{figure}
\psfig{figure=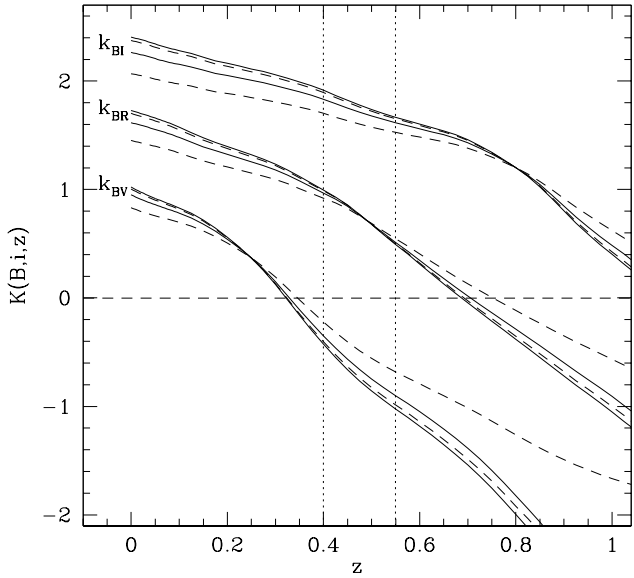,width=8.8cm}
\caption[]{The K-correction for the conversion from observed magnitudes in
  the $i=V,R$ and $I$ band, respectively, into restframe $B$
  magnitudes.  Underlying SEDs are 1-Gyr-burst models (solid lines) and
  $\tau=2$ models (dashed lines) at ages of 10 and 18\,Gyrs. The vertical
  dotted lines indicate the redshifts of the investigated clusters. }
\label{kc}
\end{figure}

The K-correction is defined as the quantity necessary in the case of
redshifted objects to convert the observed magnitude in a given filter to
the restframe magnitude.
%
%
With increasing redshift, the restframe $B$ magnitude maps successively
into the $V,R$ and $I$ bands. As a consequence, there exist different
redshift ranges in which the absolute values of the K-correction to
the B rest-frame in the
three bands are minimal. Fig.\,\ref{kc} shows these values of the
K-corrections as a function of redshift, obtained by convolving the
filter functions with different models for the galaxy SED.  We consider
synthetic SEDs constructed with the stellar population models by
Bruzual and Charlot \cite*[BC98]{BC98}, assuming a 1\,Gyr star burst as
representative of ellipticals, and a $\tau=2$ exponentially decreasing
star formation rate as representative of Sa galaxies. The adopted IMF
is Salpeter. We also consider a range of ages (i.e. 10 and 18 Gyr) for
the galaxies at $z=0$, to encompass the plausible range of
K-corrections.  The derived mean values are listed in
Table\,\ref{calib}, together with their variations due to the
different SED models.

It can be seen that for our intermediate redshift clusters the absolute
values of the K-corrections are smallest for the $V$ filter.  The
uncertainty due to the galaxy SED is on the order of a tenth of a
magnitude for our explored range of models.  Using model SEDs by
Rocca-Volmerange and Guiderdoni \cite*{RG88} (burst, cold E and Sa
models at the age of 12\,Gyrs) we derive K-corrections which fall into
approximately the same range.

To summarize, there are three sources for systematic errors in the
calibration of the surface brightness.  The transformation from HST to
UBVRI magnitudes is affected by a zero point error that we estimate to
be $\Delta m\leq0.1$\,\magn. The uncertainty in the K-correction due to
the real SED of the galaxies conveys a similar error.  According to BH
and SFD, the error due to the absorption coefficient should be lower
than this, but as stated above, the $A_i$ values derived from BH
reddenings are systematically lower than those derived with SFD
prescriptions. Note, however, that for the Coma cluster, our low
redshift reference, the differences are small, $\la 0.01$ mag. 
As a result, the luminosity evolution inferred adopting
BH reddenings is weaker by about $0.06$\,\magn.  
In the following, unless stated otherwise, we
apply the SFD absorption coefficients and the average value of the
K-correction to correct the surface brightness values.

Concerning the effective radius, we did not correct for any color
gradient, because the deviations in $r_e$ in different filters are
negligibly small compared to the error in the determination of $r_e$
itself. To transform the measured effective radii (in arcsec) into
metric units we used the \cite{M58} formula (see, e.g., Ziegler and
Bender \cite*{ZB97}):
\begin{equation}
\label{rekpc}
\frac{R_e}{\mathrm{kpc}} = 14.534 \cdot \frac{r_e}{\mathrm{arcsec}} \cdot 
\frac{q_0 z + (q_0-1) (\sqrt{1+2q_0z} -1)} {h \,q_0{}^2 (1+z)^2},
\end{equation}
where $h=H_0/(100 km s^{-1} Mpc^{-1})$.
Table\,\ref{gals} lists for each cluster the minimum, median, and
maximum values of the logarithm of $R_e$ (in
$\mathrm{kpc}$) of the ET galaxies samples. As
throughout the whole paper, the Hubble constant was taken to be
$H_0=60\,\mathrm{km\,s\me\,Mpc\me}$ and the deceleration parameter to be
$q_0=0.1$.


\section{Deriving the luminosity evolution}
\label{lumev}

\subsection{Local Kormendy relations}
\label{localKR}

\cbstart To study the evolution of distant galaxy samples it is
crucial to first understand the local comparison sample and how the
distribution of galaxies in the Kormendy diagram depends on different
selection effects. Since we want to compare galaxies in clusters, we
choose as the local reference the Coma cluster, which has a similar
richness like the distant clusters under consideration so that any
possible environmental effect is minimized \cite{JFK95a,Joerg97}. 

\begin{figure}
\psfig{figure=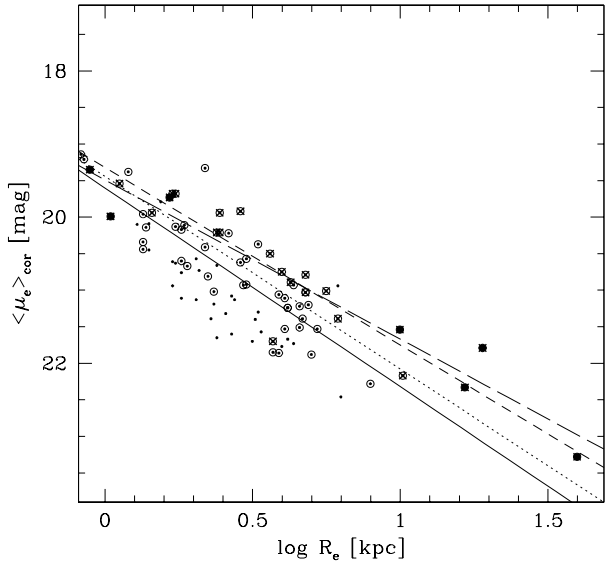,width=8.8cm}
\caption[]{The bisector fits for the JFK sample of early-type galaxies
in Coma subdivided into four $\sigma$-bins: $\sigma>0$ (dots, solid
line), $\sigma>140$ (open circles, dotted line), $\sigma>200$
(crosses, short dashed line) and $\sigma>250$\,km\,s\me\ (filled
squares, long dashed line). }
\label{jfk_sig}
\end{figure}

\begin{figure}
\psfig{figure=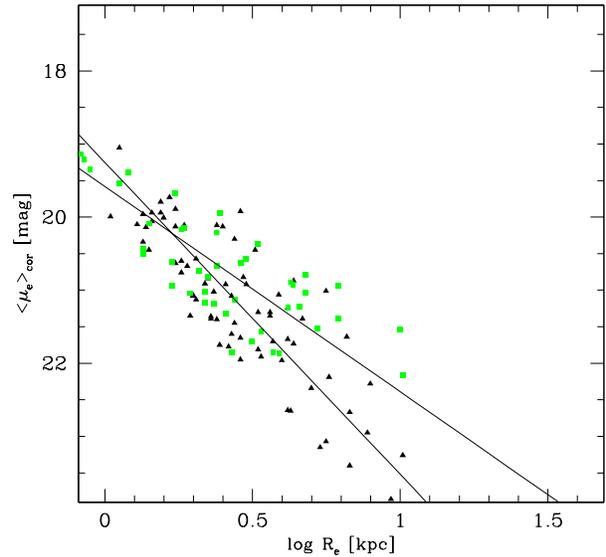,width=8.8cm}
\caption[]{The bisector fits for the JFK sample of early-type galaxies
in Coma subdivided into the morphological classes of ellipticals
(squares) and S0s (triangles). }
\label{jfk_es}
\end{figure}

As a first example we examine the sample of early-type galaxies in the
Coma cluster of J\oe rgensen et al. \cite*[JFK]{JFK95a,JFK95b}. To
transform their Gunn $r$ restframe magnitudes into Johnson $B$, we
adopt an average colour of $(B-r)=1.15$ for observed nearby E and S0
galaxies which is similar to the $(B-r)=1.02$ model colour of Fukugita
et al. \cite*{FSI95} for an E galaxy at $z=0$. In Table\,\ref{local},
we summarize the fit parameters of the Kormendy relation for various
selections we introduced to the JFK sample. The slope and zeropoint
are determined by a bisector fit to the fully corrected surface
brightness $\mathrm{\langle\mu_e\rangle_{cor}}$ as a function of the
logarithm of the effective radius given in kpc, $\log R_e$. The values
in brackets are ordinary least-square fit parameters with the
variables interchanged. The JFK sample consists of 147 early-type
galaxies of which 92 have velocity dispersion ($\sigma$)
measurements. According to the authors their sample is complete to
$r=15$ corresponding to $B=16.2$. Next we reduce the sample to those
galaxies which are within 810\arcsec of the cluster center (according
to Godwin et al. \cite*{GMP83}) corresponding to 870\,kpc which is the
field of view of the WFPC2 camera at a redshift of $z=0.4$ where most
of our analyzed clusters are located. By doing this the fit parameters
hardly get changed. Since the Kor\-mendy relation is a projection of
the Fundamental Plane, with no dependence on the velocity dispersion
$\sigma$, the coefficient $b$ of $\log R_e$ in Eq.\,(\ref{KR}) will be
different from the well-established coefficient $\tilde{b}$ in
Eq.\,(\ref{FP}). Indeed, the slope of the Kormendy relation turns out
to be different for samples covering different ranges of the velocity
dispersion. We illustrate this fact in Fig.\,\ref{jfk_sig} by
subdividing the JFK sample into four $\sigma$-bins. The slope $b$ gets
increasingly higher for samples with lower mean $\sigma$. Note, that
this effect is not caused by different magnitude cut-offs for the
various subsamples since each subsample spans almost the same range in
apparent magnitudes. But decreasing the magnitude cut-off also results
in a slight increase of the slope. At last, we investigate the effect
of subdividing the JFK early-type galaxies into ellipticals and
S0-galaxies. The morphological types are given by the authors but are
based on Dressler \cite*{Dress80b}. The distribution of the S0 and E
galaxies in the Kormendy diagram are quite distinct with a larger
slope for the S0 galaxies, see Fig.\,\ref{jfk_es}. If galaxies with
extreme values (large $R_e$ for Es, faint $\langle\mu_e\rangle$ for
S0s) are excluded the respective slopes do not deviate so much from
each other any more.

Since studies of local clusters never found significant deviations in
the distribution of galaxies in the Fundamental Plane between
different clusters
\cite{DLBDFTW87,BBF92,JFK95a} and since any dynamical evolution moves
early-type galaxies only within the Fundamental Plane \cite{CLR96} we
assume that the galaxies in the distant clusters are similarly
distributed within the FP and, therefore, also within its projection
onto the Kormendy plane. Nevertheless, we will investigate the effect
of freely determining the slope of the Kormendy
relation for all clusters, in contrast to  the practise in previous studies,
and the effect of subdividing the distant galaxies according to their
disk-to-bulge ratios.

\cbend{}

\begin{table*}
\caption[]{Local comparison samples. The second column gives the
number of galaxies of the respective sample, $a$ and $b$ are the
zeropoints and slopes of the bisector fit to the respective $B$ band
Kormendy relations. The values in brackets are $a$ and $b$ of the
least-square fits.}
\label{local}
\begin{flushleft}
\begin{tabular}{lrll}
\noalign{\smallskip}
\hline
\noalign{\smallskip}
sample & nog & a & b \\
\noalign{\smallskip}
\hline
\noalign{\smallskip}
Coma JFK: all ($12.14\le B\le16.76$) & 147 & 19.46 (19.79;18.91) & 3.46 (2.73;4.63) \\
Coma JFK: $B<16.2$    & 108 & 19.17 (19.51;18.65) & 3.59 (2.94;4.56) \\
Coma JFK: ``HST'' FOV &  74 & 19.50 (19.81;19.00) & 3.45 (2.77;4.53) \\
Coma JFK: all with $\sigma$ ($12.14\le B\le16.75$)&  92 & 19.60 (19.82;19.27) & 2.72 (2.24;3.43) \\
Coma JFK: $\sigma>140$ ($12.14\le B\le16.75$)& 62 & 19.44 (19.59;19.27) & 2.64 (2.35;3.01) \\
Coma JFK: $\sigma>200$ ($12.14\le B\le16.75$)& 24 & 19.32 (19.40;19.24) & 2.43 (2.30;2.58) \\
Coma JFK: $\sigma>250$ ($12.14\le B\le16.75$)&  7 & 19.49 (19.54;19.45) & 2.18 (2.13;2.24) \\
Coma JFK: $B<16.5$ ($68\le\sigma\le386$)& 85 & 19.51 (19.73;19.20) & 2.81 (2.35;3.45) \\
Coma JFK: $B<16.0$ ($81\le\sigma\le386$ & 64 & 19.30 (19.57;18.91) & 2.96 (2.47;3.64) \\
Coma JFK: E only      &  44 & 19.58 (19.83;19.18) & 2.81 (2.22;3.76) \\
Coma JFK: S0/SB0 only &  78 & 19.25 (19.53;18.87) & 4.27 (3.66;5.10) \\
Coma JFK: S0 with $\mathrm{\langle\-\mu_e\rangle_{cor}}<22$ &  65 & 19.44 (19.78;18.83) & 3.68 (2.81;5.23) \\
Coma JFK: E with $\log R_e<1$kpc & 42 & 19.52 (19.79;19.06) & 3.02 (2.33;4.20) \\
\hline
Coma SBD: all         &  39 & 19.80 (19.88;19.72) & 2.18 (2.03;2.35) \\
Coma SBD: no cDs      &  36 & 19.75 (19.91;19.53) & 2.33 (1.94;2.89) \\
Coma SBD: E only      &  25 & 19.70 (19.76;19.62) & 2.23 (2.11;2.37) \\
Coma SBD: S0/SB0 only &  14 & 19.86 (20.03;19.58) & 2.42 (1.90;3.23) \\
Coma JFK: same as SBD &  39 & 19.63 (19.77;19.45) & 2.46 (2.17;2.82) \\
\noalign{\smallskip}
\hline
\end{tabular}
\end{flushleft}
\end{table*}

\cbstart As a second example we take the data of Saglia et
al. \cite*[SBD]{SBD93}, which we re-calibrated and analyzed in the
same manner as we did with the distant galaxies \cite{BSZBBGH97b}.
This sample has the advantage that it ensures a uniform
fitting procedure for both the local and distant galaxies. Despite of
being morphologically selected, this sample comprises both E and
S0/SB0 galaxies but being restricted to the central part of the
cluster does not contain any post-starburst galaxy of Caldwell et
al. \cite*{CRSEB93} and, therefore, is a fair comparison to the
distant spectroscopically selected cluster samples. In
Table\,\ref{local}, we report the fit parameters to the Kormendy
relation derived for this sample, too. A slightly different slope is
found when we either include or exclude the three brightest E
galaxies. As with the JFK sample a larger slope is found for a
subsample of only S0/SB0 galaxies than for one of only ellipticals,
but the difference is marginal.
\cbend{}

We conclude that the slope of the Kormendy relation for cluster
galaxies is in the range $2.2...3.6$, with a tendency to increase from
the earlier to the later galaxy types. \cbstart In the following we
will consider both the Coma JFK and SBD samples as the local reference
to determine the evolution of the Kormendy relation. Thus, we are able
to estimate how the incompleteness of the SBD sample effects the
results. \cbend{}
 
\subsection{The method}

In order to derive the luminosity evolution of ET cluster
galaxies one has to compare the surface brightnesses as given by the
Kormendy relation of a distant cluster to a local one. Most authors
have made this comparison by choosing one slope for the Kormendy relation
for both distant and local clusters, and looking at the variation of
the surface brightness at a fixed standard effective radius of
$1\,\mathrm{kpc}$ (i.e. the variation of the zeropoint of the Kormendy
relation at $\log R_e=0$).  This corresponds to assuming that (i) the
slope of the Kormendy relation is independent of  redshift; (ii)
that its dependence on the ET galaxies selection is negligible; (iii)
that at fixed $R_e$ there is a one to one correspondence between
galaxies in the local and distant clusters.  There is no a priori
reason for these three assumptions to be valid, and indeed we have
shown in the previous section that in the Coma cluster there is a
dependence of the slope on the galaxy morphological subclass (and on
the velocity dispersion range).

\begin{figure}[htb]
\psfig{figure=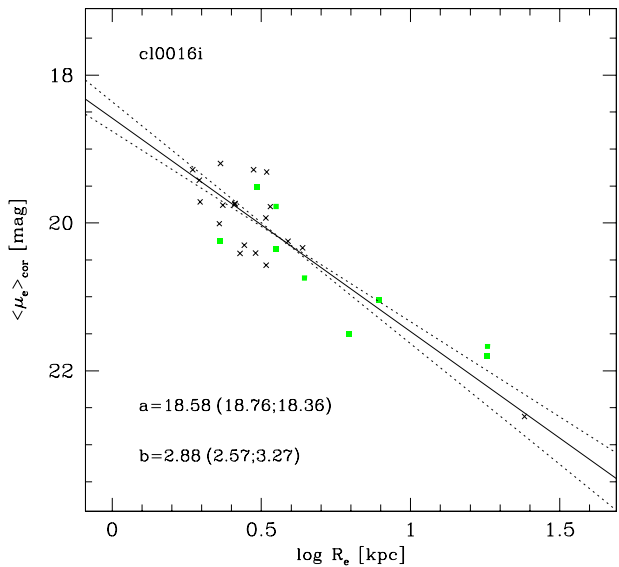,width=8.8cm}
\caption[]{The Kormendy relation of \textit{cl0016i}-galaxies. The solid
  line represents the bisector fit, the dotted lines are the lsq fits
  with variables interchanged. $a$ is the zeropoint, $b$ the slope of
  the bisector fit (values in brackets correspond to the lsq fits). The squares
  are galaxies with $d/b>0.2$. }
\label{bisect}
\end{figure}

In order to evaluate the evolution of the surface brightness of ET
galaxies in clusters taking into account a possible variation of the
slope of the Kormendy relation with redshift we proceed in two ways:
\par\noindent
(1) fitting independently the Kormendy relations for the various high
$z$ clusters and comparing them to the relation derived for Coma;
\par\noindent
(2) imposing a fixed slope for the Kormendy relation in all the
clusters, and exploring the derived luminosity evolution for a range
of values for this slope. 
 
\subsection{The Kormendy relations in the distant clusters}
\label{distKR}

Before we describe how the Kormendy relation depends on
redshift, we first make sure that there exists a correlation between
$\mathrm{\langle\-\mu_e\rangle_{cor}}$ and $\log\-R_e$ for the distant
galaxy samples in a statistical sense.  We performed a Spearman's rank
analysis and find that all
samples with more than ten galaxies show indeed a correlation on the
99\% probability level.

\begin{figure}[htb]
\psfig{figure=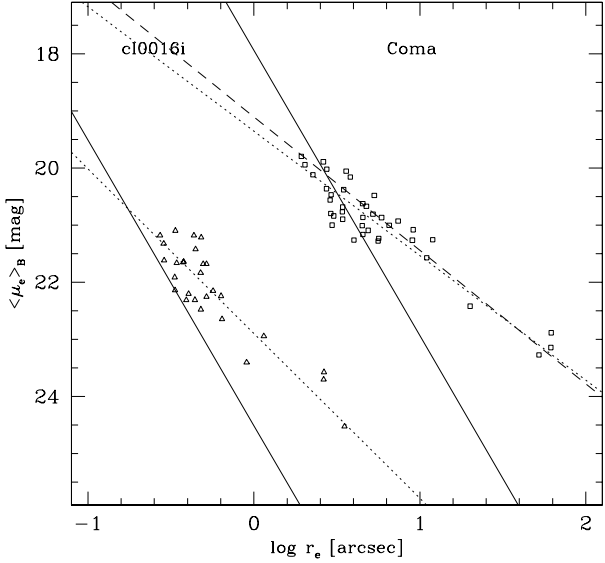,width=8.8cm}
\caption[]{The (observed) Kormendy relation (dotted lines) for the distant
  cluster sample \textit{cl0016i} (triangles) compared with the local sample
  (squares). The solid line represents the magnitude limit, which is
  shifted to the Coma\,SBD sample according to the calculated distance
  modulus and the expected luminosity evolution. The dashed line is the
  fit to this reduced sample. }
\label{comp}
\end{figure}

For each distant cluster we determine the slope and zero points of the
Kormendy relations for all ET galaxies (i.e., no \ea) by performing a
bisector fit, as done for the Coma cluster. The results for each cluster
are listed in Table\,\ref{bct}.  Fig.\,\ref{bisect} shows the fit for
\textit{cl0016i} as an example. The slopes of the Kormendy relations
for the distant clusters scatter between 2.2 and 3.5, which is within
the range of the quoted local slopes (see Table\,\ref{local}). This
indicates that the slope of the Kormendy relation does not change
significantly with redshift. Since the effective radii of the galaxies
in the distant clusters span a similar and wide range
($2.5\la\-R_e/\mathrm{kpc}\la20$) (see Table\,\ref{gals}), this implies
that on the average the stellar populations of smaller ellipticals have
not evolved from $z=0.6$ until today in a markedly different way with
respect to those of larger ellipticals. A natural explanation for this
is that the mean ages of the stars in small galaxies are not very
different from those in large galaxies, implying a rather old age and
high formation redshifts for their stellar populations, independent of
the size of the early-type galaxies.

The differences in the zeropoints $a$ in Table\,\ref{bct} with respect
to the same value for Coma reflect the surface brightness evolution of
galaxies with $R_e = 1$kpc, having adopted the Kormendy relation which
best fits the data for the individual clusters. It can be noticed
that for the four clusters at $z\approx0.4$ we find a substantial
scatter of these zeropoints.  This stems from having considered the
zeropoint at $R_e=1$ kpc. To describe the {\it global} evolution of cluster
ellipticals, it is more meaningful to compare the surface brightnesses
at the median value of the effective radii distribution ($\langle \log
R_e \rangle$).  Indeed the scatter of the surface brightness at
$\langle \log R_e \rangle$ for the 4 clusters at $z\approx 0.4$ is
substantially reduced (see column 6 of Table 4).

\begin{figure}[tb]
\psfig{figure=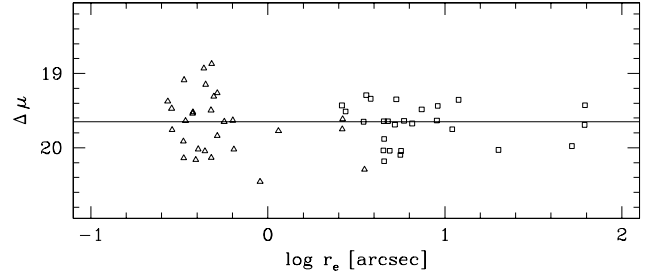,width=8.8cm}
\caption[]{The distribution of the residual surface brightnesses
  $\Delta \mathrm{\mu}_i$ of \textit{cl0016i} (triangles) and Coma SBD
  (squares) around their median value (solid line). The $\Delta
  \mathrm{\mu}_i$ of \textit{cl0016i} were calculated here with the
  slope $b$ fixed to the Coma SBD value of $2.18$ instead of $2.88$ of
  the free bisector fit and are shown already reduced by the evolution
  $\Delta M_{\mathrm{fixed}}$.  }
\label{delsb}
\end{figure}

Selection effects must be taken into account before these values can be
used to derive the luminosity evolution. Obviously for the Coma cluster
the distribution of effective radii extends down to much lower values
than those reached in the distant samples. In order to apply similar
selections for both the distant and the local samples, we cut off the
Coma samples at a suitable magnitude limit. In this way we
also minimize to first order the bias induced by the galaxy
distribution in $\sigma$ (see Fig. \ref{jfk_sig}). To determine the
magnitude cut-off in the distant clusters introduced by the selection
of the ET galaxies we go back to Eqs. (\ref{sbhst}) and
(\ref{sbrest}). The maximum of the total magnitudes of all galaxies $i$
in a given sample represents the respective magnitude limit:
\begin{equation}
\label{blim}
B_{\mathrm{lim}} = \max_i \left( \mathrm{\langle \mu_e\rangle} _{B,i} - 
5 \log (r_{e,i}) - 2.5 \log (2 \pi) \right)
\end{equation}
This limit is then applied to the Coma sample taking into account the
difference in distance modulus.  This procedure may not be correct, due
to the luminosity evolution of individual galaxies. For example,
passive evolution will force the fainter galaxies in the distant
clusters to fade below the magnitude limit determined the way just
described. To correct for this, we have to assume a luminosity
evolution for the galaxies at the faint end of the distribution,
$\Delta M_{B,\mathrm{lim}}$. As a first attempt we take a fading of
$\Delta M_{B,\mathrm{lim}} = 0.5$ and $0.66$ \magn\ for the clusters at
$z=0.4$ and $0.55$, respectively. These values are close to what is
expected from Bruzual and Charlot models for the passive evolution of
old stellar populations, and in the following will be referred to as
initial set of parameters.  Different values for the $\Delta
M_{B,\mathrm{lim}}$ are explored later.  Fig.\,\ref{comp} illustrates
how the magnitude cut-off is implemented to estimate the luminosity
evolution for \nnes.

To summarize, for each image of each cluster we construct the median
$R_e$, calculate the surface brightness at this median $R_e$ from the
Kormendy relation found for the ET cluster members in the specific
image, and compare it to the surface brightness coming from the
Kormendy relation constructed for the subsample of Coma ET galaxies
brighter than the appropriate cut-off magnitude, evaluated at $\langle
\log R_e \rangle$. The results are listed in columns 6 to 8 of
Table\,\ref{bct}. The luminosity evolution determined with this {\it
free slope} 
approach is referred to as $\Delta M_{\mathrm{free}}$.

The last columns in Table\,\ref{bct} list the estimate of the evolution
of the surface brightness with the second method, i.e. enforcing the
same slope for the Kormendy relation in both the local and the distant
clusters. We initially choose $b=2.18 (3.46)$, which is appropriate for the
Coma SBD (JFK) sample. The residual surface brightness of a galaxy $i$ in a
specific image is defined as:
\begin{equation}
\Delta \mathrm{\mu}_i = 
\langle \mathrm{\mu}_e \rangle_{\mathrm{cor},i}-b\log R_{e,i}
\end{equation}
The analogous value for Coma is derived considering only ET galaxies
brighter than the appropriate cut-off.  We prefer median values instead
of means as a robust procedure to take care of outliers. The luminosity
evolution determined with this approach is referred to as $\Delta
M_{\mathrm{fixed}}$. Figure\,\ref{delsb} describes the method for one
cluster. Note that the residuals are not equally distributed around the
fit: for this case, the actual slope of the Kormendy relation is
$2.88$, while a fixed slope of $2.18$ has been adopted to compute the
median surface brightness. Note that the difference between the two
numbers is not significant in the light of Table \ref{local}.

\begin{table*}[htb]
\caption[]{\cbstart Results for the initial set of parameters (SFD extinction,
$H_0=60\,\mathrm{km\,s\me\,Mpc\me}, q_0=0.1$, $\Delta
M_{B,\mathrm{lim}}=0.5 (0.66)\magn$ at $z=0.4 (0.55)$).
Col.\,3: number of galaxies used for the fits, Cols.\,4. \& 5.:
zeropoint and slope of the bisector fit, Col.\,6: surface brightness
at the median effective radius of the sample calculated according to
$a$ and $b$, Col.\,7: the same for the Coma sample (SBD or JFK) which was
reduced by the relevant magnitude cut-off, Col.\,8: difference between
$\mu(\langle R_e\rangle)$ and $\mu(\langle R_e\rangle)_c$. Col.\,9:
median value of the residual surface brightnesses
$\Delta\mathrm{\mu}_i$, Col.\,10: the same for the reduced Coma sample,
Col.\,11: difference between $\langle\Delta\mu\rangle$ and
$\langle\Delta\mu_c\rangle$. Cluster samples written in italic are not
used in the statistical analysis because they contain too low a number of
galaxies ($<10$). \cbend{} }
\label{bct}
\begin{flushleft}
\begin{tabular}{llr|llllr|llr}
\noalign{\smallskip}
\hline
\noalign{\smallskip}
cluster & z & nog & a & b & $\mu(\langle R_e\rangle)$ &
$\mu(\langle R_e\rangle)_c$ & $\Delta M_{\mathrm{free}}$ & 
$\langle\Delta\mathrm{\mu}\rangle$ & $\langle\Delta\mathrm{\mu_{c}}\rangle$ &
$\Delta\-M_{\mathrm{fixed}}$
\\
\noalign{\smallskip}
\hline
\noalign{\smallskip}
ComaSBD & 0.024 & 39 & 19.80 & 2.18 & 20.69 & 20.67 &    0.02 & 19.75 & 19.75 &    0.00  \\
\textit{  a370v} & 0.375 &  9 & 17.40 & 4.52 & 20.25 & 20.93 & $-$\textit{0.68} & 18.86 & 19.64 & $-$\textit{0.77} \\
  a370r & 0.375 & 17 & 18.71 & 2.52 & 20.39 & 21.02 & $-$0.64 & 18.95 & 19.64 & $-$0.69  \\
\textit{  a370i} & 0.375 &  9 & 18.14 & 3.81 & 20.26 & 20.78 & $-$\textit{0.52} & 19.11 & 19.64 & $-$\textit{0.53} \\
cl1447r & 0.389 & 31 & 18.77 & 3.51 & 20.24 & 20.70 & $-$0.46 & 19.30 & 19.75 & $-$0.44  \\
\textit{cl0939v} & 0.407 &  8 & 19.18 & 2.70 & 20.59 & 20.88 & $-$\textit{0.28} & 19.58 & 19.69 & $-$\textit{0.12} \\
cl0939r & 0.407 & 26 & 19.34 & 2.58 & 20.43 & 20.73 & $-$0.29 & 19.31 & 19.75 & $-$0.44  \\
\textit{cl0939i} & 0.407 &  6 & 19.92 & 1.27 & 20.53 & 20.80 & $-$\textit{0.27} & 19.56 & 19.69 & $-$\textit{0.13} \\
cl0303r & 0.416 & 24 & 19.06 & 2.88 & 20.58 & 20.86 & $-$0.28 & 19.42 & 19.67 & $-$0.25  \\
cl0016v & 0.550 & 30 & 18.17 & 3.23 & 19.75 & 20.66 & $-$0.91 & 18.75 & 19.64 & $-$0.89  \\
cl0016i & 0.550 & 28 & 18.58 & 2.88 & 20.07 & 20.80 & $-$0.73 & 18.96 & 19.65 & $-$0.69  \\
\noalign{\smallskip}
\hline
\noalign{\smallskip}
ComaJFK & 0.024 &147 & 19.46 & 3.46 & 20.97 & 20.97 &    0.01 & 20.04 & 20.04 &    0.00 \\
\textit{  a370v} & 0.375 &  9 & 17.40 & 4.52 & 20.25 & 20.89 & $-$\textit{0.64} & 18.86 & 19.54 & $-$\textit{0.67} \\
  a370r & 0.375 & 17 & 18.71 & 2.52 & 20.39 & 21.20 & $-$0.81 & 18.95 & 19.67 & $-$0.72 \\
\textit{  a370i} & 0.375 &  9 & 18.14 & 3.81 & 20.26 & 20.81 & $-$\textit{0.55} & 19.11 & 19.62 & $-$\textit{0.51} \\
cl1447r & 0.389 & 31 & 18.77 & 3.51 & 20.24 & 20.82 & $-$0.59 & 19.30 & 19.96 & $-$0.66 \\
\textit{cl0939v} & 0.407 &  8 & 19.18 & 2.70 & 20.59 & 21.07 & $-$\textit{0.48} & 19.58 & 19.85 & $-$\textit{0.28} \\
cl0939r & 0.407 & 26 & 19.34 & 2.58 & 20.43 & 20.92 & $-$0.49 & 19.31 & 20.04 & $-$0.72 \\
\textit{cl0939i} & 0.407 &  6 & 19.92 & 1.27 & 20.53 & 20.94 & $-$\textit{0.41} & 19.56 & 19.89 & $-$\textit{0.33} \\
cl0303r & 0.416 & 24 & 19.06 & 2.88 & 20.58 & 21.03 & $-$0.45 & 19.42 & 19.80 & $-$0.38 \\
cl0016v & 0.550 & 30 & 18.17 & 3.23 & 19.75 & 20.60 & $-$0.86 & 18.75 & 19.67 & $-$0.92 \\
cl0016i & 0.550 & 28 & 18.58 & 2.88 & 20.07 & 20.88 & $-$0.81 & 18.96 & 19.78 & $-$0.81 \\
\noalign{\smallskip}
\hline
\end{tabular}
\end{flushleft}
\end{table*}

\begin{figure*}[htb]
\psfig{figure=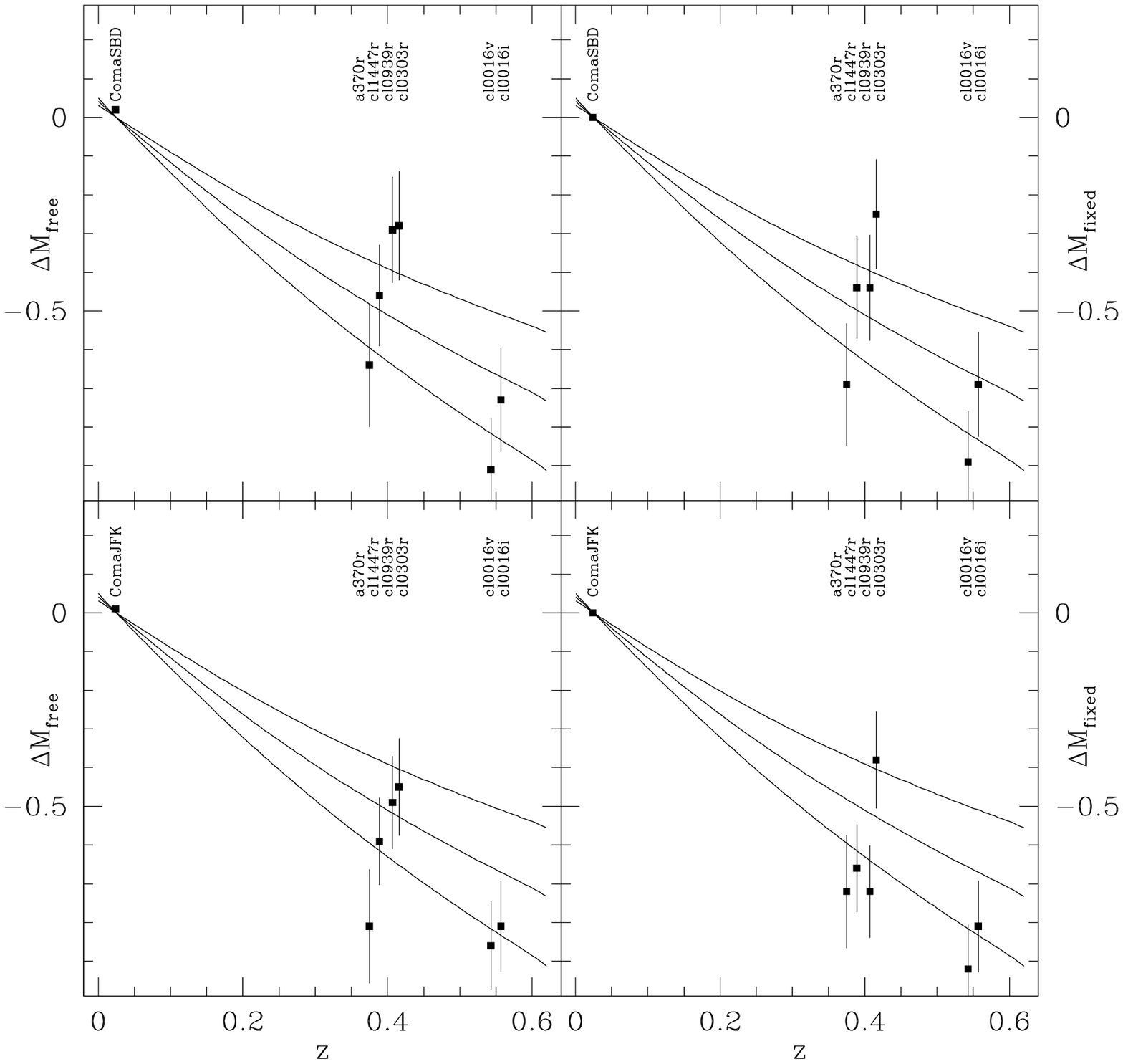,width=18cm}
\caption[]{Luminosity evolution derived from the Kormendy relations
  for the initial set of parameters (see text). \cbstart Top panels:
  local reference sample: Coma SBD, bottom panels: local reference
  sample: Coma JFK. Left panels: free bisector fit method, right
  panels: fixed slope fit method (see text for explanations). \cbend{}
  Solid curves: expected evolution from a BC98 model with $z_f=4$ and
  IMF slope $x=2.35, 1.35$ (Salpeter) and $0.35$ (top to bottom). The
  points for the V and I images of \textit{Cl\,0016} are shifted in
  $z$ for better visibility. The errorbars are the quadratic sum of
  the mean $\sigma_c$ and the standard deviation of the Coma sample. }
\label{bcf}
\end{figure*}

The results of the two methods are visualized in Fig.\,\ref{bcf} for
the images in which more than 10 galaxies could be used (i.e. the $R$
images of \adsn, \ndnd, \nndn\ and \textit{Cl\,1447$+$26}, and the $V$
and $I$ images of \nnes) and compared to both Coma samples (SBD and
JFK). The errors are computed in the following way: we first
calculate the standard deviation of the residuals of all galaxies
($n_g$) around the bisector fit to the $\mathrm{\langle
\mu_e\rangle_{cor}}$--$\log(R_e)$ data for each cluster individually:
$\sigma_g = \sqrt{\sum_{i=1}^{n_g} (x_i - \bar{x})^2/(n_g-1)}$ with $x
= \mathrm{\langle \mu_e\rangle_{cor}} - b \cdot \log(R_e) - a$ and
$\bar{x} = \sum_{i=1}^{n_g} x_i / n_g$.
%
%
The average observed scatter is then taken to be the combined standard
deviation of all clusters ($n_c$):
$\sigma_a = \sqrt{\sum_{i=1}^{n_c} \sigma_{g,i}^2/n_c}$.
%
%
The error for each cluster is then: $\sigma_c = \sigma_a / \sqrt{n_g}$.
%
%
Note that the scatter in the Kormendy relation arises mainly from 
neglecting the velocity dispersion of the tight Fundamental Plane and is
little augmented by the measurement errors. See Sect. \ref{model} for a
discussion of the errors induced by the K-corrections.

\cbstart The overall redshift evolution of the surface brightness of
ET galaxies derived with the two methods and compared to the two local
reference samples is quite similar. Distinctive differences between
the same individual samples in the 4 panels of Fig.\,\ref{bcf} are not
significant given the large errors in the single data points. 
 Because both our methods rely on median values
the incompleteness of the Coma SBD sample does not have a systematic
effect on the derived evolution. The overall slightly higher values of
$\Delta M$ for the JFK sample are rather the result of the
transformation from Gunn $r$ to Johnson $B$ magnitudes.  \cbend{} The
predictions of passive evolution models are also shown in
Fig.\,\ref{bcf}. These are BC98 models calculated for a 1-Gyr-burst
population of solar metallicity forming at $z=4$ 
($t_{\mathrm{gal}}=12$\,Gyr) and IMF slopes of
$x=2.35, 1.35$ (Salpeter) and $0.35$, respectively.  The data points
for the cluster samples are compatible with the considered models
within the 1-$\sigma$-error in all four panels.

\section{Effects of the assumptions}
\label{para}

The luminosity evolution derived in the previous section using method
(1) or (2) depends on various assumptions. In the following we 
explore the effects of the assumptions used for method (2):
\par\noindent
(i) the value of the slope of the Kormendy relation adopted for
all the clusters, 
\par\noindent
(ii) the value of the $\Delta M_{B,\mathrm{lim}}$ parameter,
\par
\begin{figure}
\psfig{figure=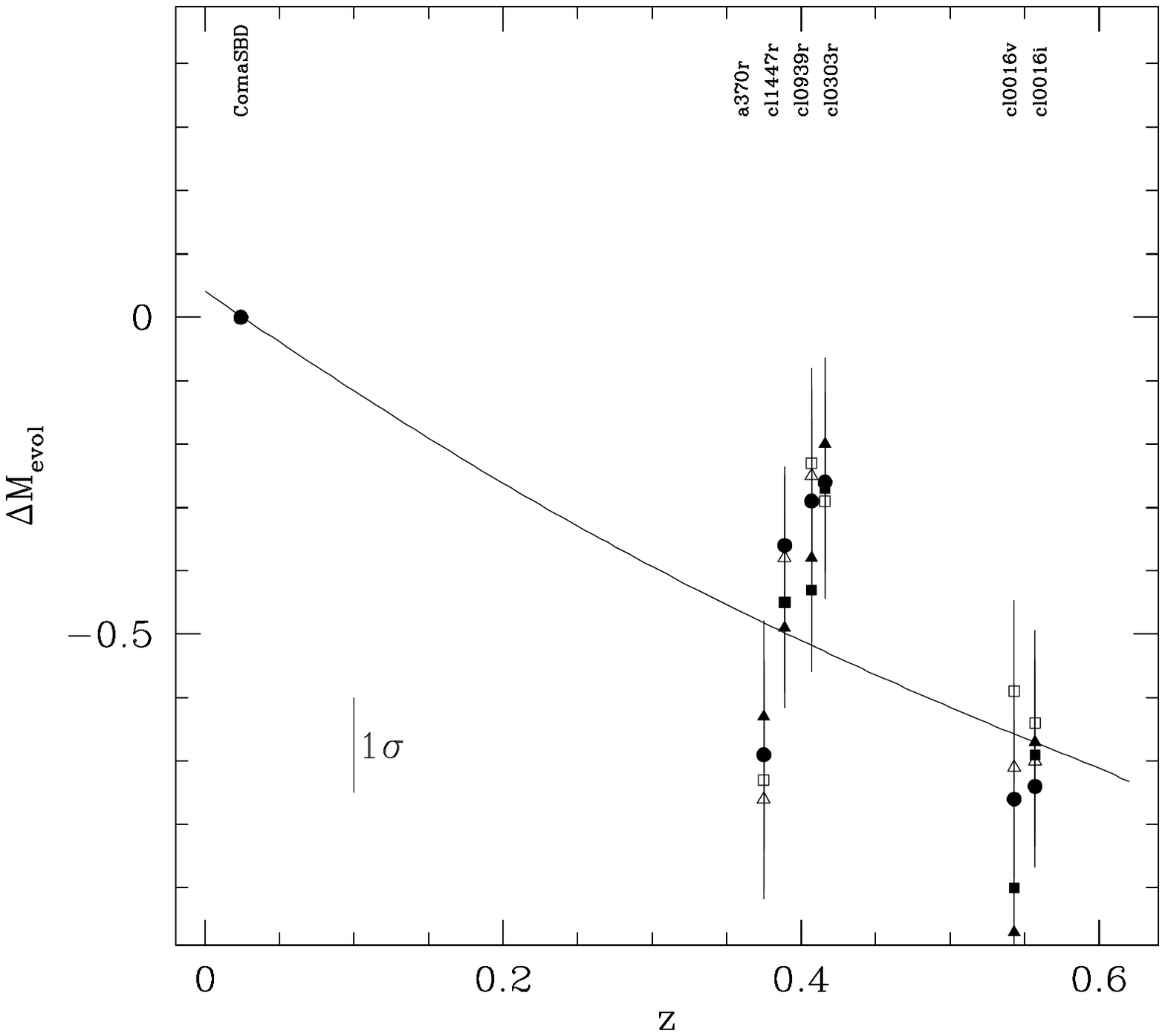,width=8.8cm}
\caption[]{Luminosity evolution derived from the Kormendy relations assuming
  different slopes: filled triangle: $b=1.9$, filled square: $2.3$,
  filled circle: $2.7$, open triangle: $3.1$, open square: $3.5$. The
  errorbar in the lower left corner represents the quadratic sum of the
  mean $\sigma_c$ and the standard deviation of the Coma sample. The
  solid curve is the BC98 model with $z_f=4$ and Salpeter IMF. }
\label{slopes}
\end{figure}

\begin{figure}
\psfig{figure=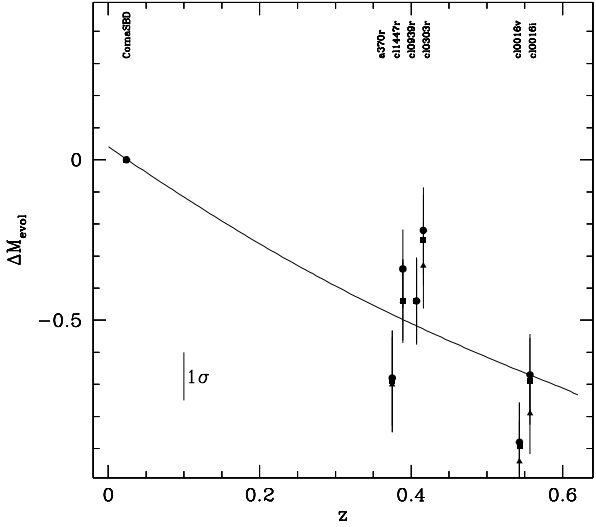,width=8.8cm}
\caption[]{Luminosity evolution derived from the Kormendy relations
assuming different evolution a priori to find the magnitude cut-off
for the Coma sample: circles: $\Delta M_{B,lim}=0$, squares: $\Delta
M_{B,lim}$ according to our 1\,Gyr burst passive evolution model,
triangles: $\Delta M_{B,lim}$ twice as high.}
\label{devol}
\end{figure}

\begin{figure}
\psfig{figure=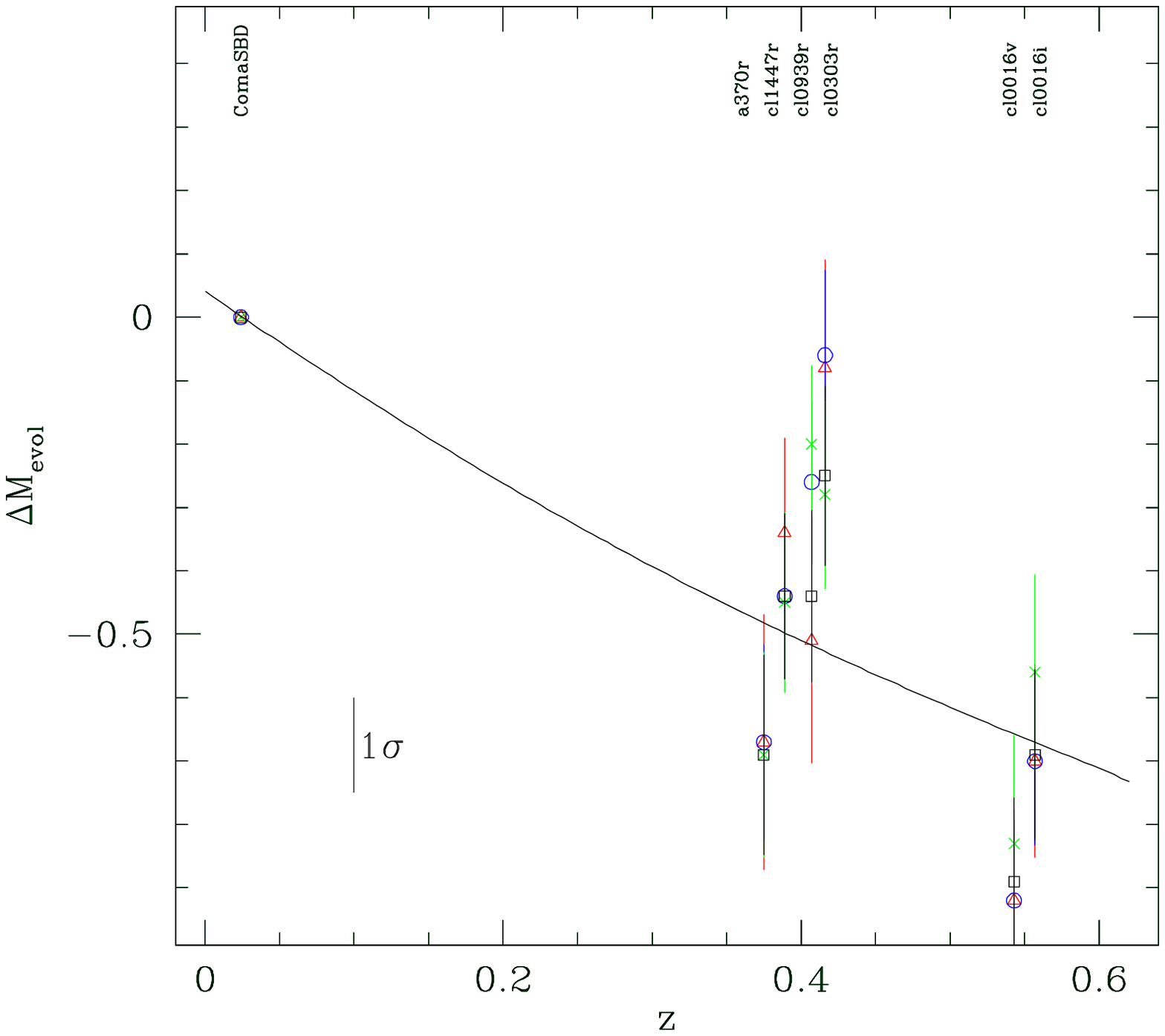,width=8.8cm}
\caption[]{Luminosity evolution derived from the Kormendy relations for
  samples with different galaxy types: squares: original samples of Es,
  S0s and Sas, triangles: ``no S0/Sas'' (only galaxies with
  $d/b\leq0.2$), crosses: ``only S0/Sas'' (only galaxies with
  $d/b>0.2$), circles: original samples with a few \ea s added. }
\label{evpop}
\end{figure}

\noindent
(iii) the selection criteria for the early type galaxies in the various
clusters.  \\
Similar tests have been performed for method (1). The detected
systematic effects are of similar size. In Fig. \ref{slopes} to \ref{add} the
luminosity is labelled $\Delta M_{\mathrm{evol}}$.

As discussed before, the slope of the Kormendy relation depends on the
range of the velocity dispersions, and on the morphological selection
criteria. It is therefore appropriate to explore the effect on the
luminosity evolution by adopting different slopes for the Kormendy
relation in our clusters. \cbstart We repeat the determination of
$\Delta M_{\mathrm{fixed}}$ as in the previous section (with Coma\,SBD
as the local sample), assuming $b=1.9, 2.3, 2.7, 3.1$, and $3.5$,
respectively. \cbend{} These values span the range of plausible slopes
observed in local samples, see Table\,\ref{local}.  The results are
shown in Fig.\,\ref{slopes}, where the different symbols refer to the
different slopes, and the line is the expected passive evolution
computed for the BC98 1-Gyr-burst model (only Salpeter IMF). \cbstart
The values of the luminosity evolution obtained with these different
slopes scatter around those obtained with the slope closest to the
free bisector fit slope. \cbend{} The differences in the derived
evolution arise from the fact that the distant galaxy samples are not
fitted by their appropriate slope.  This can be seen in the
distribution of the residual surface brightnesses of \textit{cl0016i}
in Fig.\,\ref{delsb}.

Another important assumption made in the derivation of the luminosity
evolution concerns the value of $\Delta M_{B,\mathrm{lim}}$ applied to
the ET galaxies in Coma.  It is worth mentioning that the application
of a $\Delta M_{B,\mathrm{lim}}$ does not influence the distant galaxy
samples at all. It only adds or subtracts some Coma ellipticals at the
faint end of their magnitude distribution, thus affecting $\langle
\Delta \mathrm{\mu_c}\rangle$, which is compared to the not changing
median value of the distant samples. To test the influence of this
parameter on the derived luminosity evolution we repeat the
determination of $\Delta M_{\mathrm{fixed}}$ for two more values of the
$\Delta M_{B,\mathrm{lim}}$ parameter. Assuming no evolution at all
(i.e.  $\Delta M_{B,\mathrm{lim}}=0$), the effect on the derived
luminosity evolution is rather small, as can be seen in
Fig.\,\ref{devol}. On the other extreme side, we take twice the value
expected from passive evolution models: $\Delta M_{B,\mathrm{lim}}=1.0$
\magn\ for the clusters at $z=0.4$ and $\Delta M_{B,\mathrm{lim}}=1.32$
\magn\ for \nnes\ at $z=0.55$.  The magnitude cut-off for Coma 
now implies that virtually the whole SBD sample is used for the comparison
with the distant clusters. As a result, $\langle \Delta
\mathrm{\mu_c}\rangle$ for the selected galaxies in Coma stays constant
at a high level, and the derived evolution is increased with respect to
the case discussed in the previous section. However, even with this
extreme assumption about the luminosity evolution of the fainter
galaxies, the derived surface brightness evolution is still within the
1-$\sigma_c$ error (see Fig.\,\ref{devol}).

Finally, we study the influence of our galaxy selection criteria.  Up
to now, the considered samples comprised all galaxies whose spectral
energy distributions resemble those of early-type galaxies, regardless
of their morphology. To exclude any contribution to the integrated
light by a young stellar population which might reside in a disk
component we reduce our galaxy sample for each cluster now to those
galaxies that have a disk-to-bulge ratio $d/b\leq0.2$. These subsamples
should contain neither lenticular (S0) galaxies nor extreme disky
ellipticals. The original galaxy samples are reduced by about a
factor of two by this selection (see Table\,\ref{gals}).  In spite of
the appreciably lower number of objects, the newly determined values
of the luminosity evolution are within the 1-$\sigma_c$ error with
respect to those previously obtained for the galaxy samples including
all early-type galaxies. In Fig.\,\ref{evpop} we also show the results
for the subsamples selected by $d/b>0.2$. It can be seen that there is
no trend towards weaker or stronger evolution. This means that the
original samples are not contaminated by galaxies with a disk
population substantially younger than the global average. If the
galaxies with high $d/b$-ratios in the distant clusters are really
comparable to those classified as lenticular in the nearby Universe,
then, there exists a number of S0 galaxies in clusters even at
intermediate redshifts that have disks of mainly old stars. This is
consistent with the local Fundamental Plane relation that shows no
offset between E and S0 galaxies \cite{JFK96}.

The values of the derived luminosity evolution does not change
significantly, too, if we add a few \ea\ galaxies to the original
sample of early-type galaxies (see Fig.\,\ref{evpop}). The fraction of
\ea s makes up about 10 to 20\% of the resulting samples
(Table\,\ref{gals}). This could represent a lower limit to the global
fraction of \ea\ galaxies, 
because we look at the cores of clusters, where
\ea s may be less frequent than in the outer parts \cite[and references
therein]{BR96,BVR97}.  Most of the (spectroscopically classified) \ea
s have high $d/b$-ratios. This points to spiral galaxies as the
progenitors of \ea s and not ellipticals having had a small starburst
\cite{WKK94,BBHSZ96,Wirth97}. Nevertheless, the contamination of a sample of
early-type galaxies by a small fraction of \ea s does not change the
Kormendy relation and the observed scatter is only slightly increased.

\section{Modelling the luminosity evolution}
\label{model}

\cbstart We now investigate which evolutionary stellar population
models can fit the data within their errors. For the comparison
between models and observations we arbitrarily choose one specific set
of data points, but take into account the systematic errors arising
from this particular choice. We choose the luminosity evolution as
given by $\Delta M_{\mathrm{fixed}}$ with respect to the Coma SBD
sample, because having no post-starburst galaxies its selection is
closest to the one applied for the distant clusters. We consider the
derivation of $\Delta M_{\mathrm{fixed}}$ for our initial set of
parameters with one exception: instead of applying the SFD absorptions
we take the mean of SFD and BH values and, therefore, introduce
another systematic error in our error budget. This error budget is
summarized in Table\,\ref{sigtot}. It comprises the average
statistical error for a distant sample, $\sigma_{\mathrm{stat}}$,
which is the quadratic sum of the mean $\sigma_c$ and the standard
deviation of the Coma sample:
$\sigma_{\mathrm{stat}}=\sqrt{\langle\sigma_c\rangle^2+\sigma_{\mathrm{Coma}}^2}\approx0.14\,\magn$
and several systematic errors, which must be added linearly. There are
three errors arising from the calibration of the magnitudes (see
Sect.\,\ref{seccal}): determination of the zeropoint and color
transformation, $\sigma_{\mathrm{zp}}\approx0.07\,\mathrm{mag}$,
K-correction, $\sigma_{\mathrm{kc}}\approx0.04\,\mathrm{mag}$ (for the
$R$ images of the $z=0.4$ clusters considered here, see
Table\,\ref{calib}), and extinction,
$\sigma_{\mathrm{ex}}\approx0.03\,\mathrm{mag}$ (half the average
difference between SFD and BH values). Another systematic error is
introduced by the selection of a given fixed slope for the Kormendy
relation for all clusters. From the distribution of the derived values
of $\Delta M_{\mathrm{fixed}}$ for different slopes $b$ we estimate
this error to be $\sigma_{\mathrm{fs}}\approx0.04\,\mathrm{mag}$ (see
Section\,\ref{para}). Therefore, the systematic errors add up to even
a higher value than the average statistical error (see
Table\,\ref{sigtot}). \cbend{}

\begin{table}[htb]
\caption[]{Errors introduced in the derivation of the luminosity evolution
arising from the observed scatter, the determination of zeropoint,
K-correction and extinction, and the choice of a given fixed slope. }
\label{sigtot}
\begin{flushleft}
\begin{tabular}{l|llll|l|l}
\noalign{\smallskip}
\hline
\noalign{\smallskip}
$\sigma_{\mathrm{stat}}$ & $\sigma_{\mathrm{zp}}$ & $\sigma_{\mathrm{kc}}$ &
$\sigma_{\mathrm{ex}}$ & $\sigma_{\mathrm{fs}}$ & $\sigma_{\mathrm{syst}}$ &
$\sigma_{\mathrm{tot}}$ \\
\noalign{\smallskip}
\hline
\noalign{\smallskip}
 0.14 & 0.07 & 0.04 & 0.03 & 0.04 & 0.18 & 0.32 \\
\noalign{\smallskip}
\hline
\end{tabular}
\end{flushleft}
\end{table}

The total error is quite large, as it amounts already to half the
expected value of the luminosity evolution in passive evolution models
at $z=0.4$ (see below). Seen together with the scatter of
the data points of different clusters at the same redshift it is
obvious that there is not a single evolutionary model favoured but a
broad range of models can fit the data. In the following we explore the
different allowed star formation histories using BC98 models.

\begin{figure}
\psfig{figure=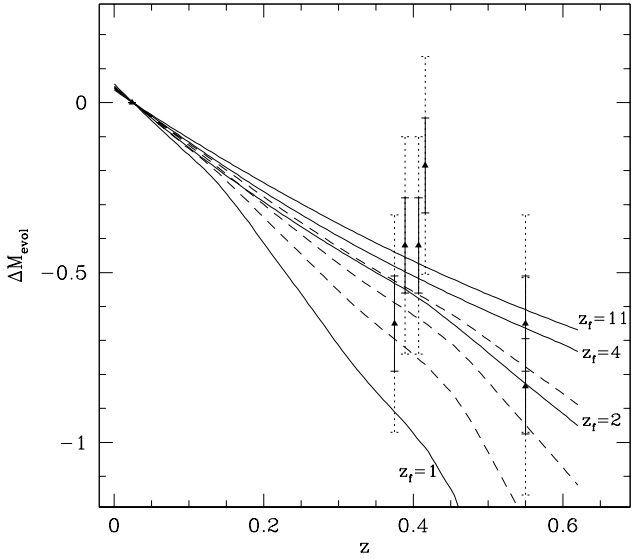,width=8.8cm}
\caption[]{Passive evolution 1-Gyr-burst models with formation epochs
  $z_f=1,2,4$ and $11$ (Salpeter IMF) superposed as solid lines onto
  the observed data with their statistical (solid errorbars) and
  systematic errors (dashed errorbars). Dashed lines are 1:1, 1:4 and
  1:10 combinations (bottom to top) between $z_f=1$ and $z_f=4$ models
  (see text). }
\label{zf}
\end{figure}

If we first confine to pure passive evolution models with an initial
1-Gyr star burst, it can be seen in Fig.\,\ref{zf} (solid lines) that
the formation epoch could be at any redshift larger than $\approx
2$, corresponding to epochs greater than $\approx 10$\,Gyrs ago. 
A more recent formation would yield a too large luminosity
evolution.  On the other hand our measured evolution reflects the
behaviour of the {\it average} properties of ET galaxies in the
clusters. \cbstart {\it Individual} galaxies could well lie away from
the average relation without violating our previous finding that the
slope of the Kormendy relation does not change significantly with $z$
given the large uncertainties in the local slope (see
Section\,\ref{distKR}). Therefore, differences in the formation
redshift between individual galaxies and the majority of the whole
cluster sample are possible even within the framework of passive
evolution. \cbend{} The dashed lines in Fig.\,\ref{zf} show the effect
on the average Kormendy relation of assuming that some fraction of the
ET galaxies in the clusters formed at lower redshifts. It can be seen
that if we assume that 10\% of the observed galaxies had formed only
at $z_f=1$ (with the remaining 90\% at $z_f=4$), and both subsamples
were equally distributed in $R_e$, the effect on the average Kormendy
relation would be small.  Only the case of 50\% galaxies formed at
$z_f$=4 and 50\% galaxies at 1 is highly disfavoured, when considering
the highest redshift cluster.

\begin{figure}
\psfig{figure=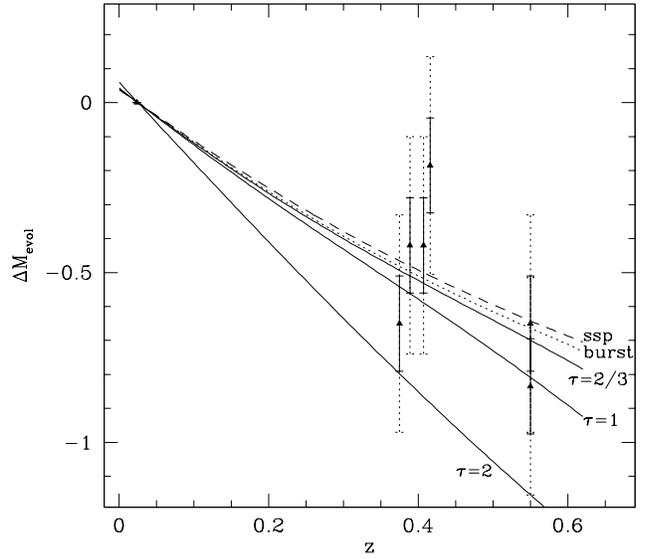,width=8.8cm}
\caption[]{Different evolutionary BC98 models with formation epoch $z_f=4$
  superposed as lines onto the observed data with their
  statistical (solid errorbars) and systematic errors (dashed
  errorbars). From top to bottom: SSP, 1-Gyr burst, $\tau=2/3, 1$ and $2$
  models. }
\label{tau}
\end{figure}

In the same way, models different from a pure burst are also compatible
with the data. Here, we investigate BC98 $\tau$-models which have
exponentially decreasing star formation rates. As an example we show in
Fig.\,\ref{tau} the predictions from models with timescales
$\tau=2/3,1$ and $2 \mathrm{Gyrs}$, all for a formation redshift
$z_f=4$ and Salpeter IMF. The dashed and dotted lines are the BC98
models for an instantaneous burst and a 1-Gyr burst, and are shown for
comparison.  Our spectrophotometric classification (see
Sect.\,\ref{sample}) assigns model galaxies with $\tau<2$ to the
family of early-type galaxies.  The $\tau$ model with star formation
timescales shorter than 1 Gyr still gives a nice representation of the
data. Thus models with currently ongoing star formation on a low level
can not be strictly ruled out, although most of the stars in the ET
galaxies must have formed at large redshift. The value of the limit on
$\tau$ depends on the assumed IMF exponent, and longer star formation
timescales would be allowed in combination with steeper IMF slopes.
However, other arguments tend to disfavour IMF slopes steeper than Salpeter
in ET galaxies (Arimoto \& Yoshii 1987, Matteucci 1994, Thomas,
Greggio \& Bender 1999).

\begin{figure}
\psfig{figure=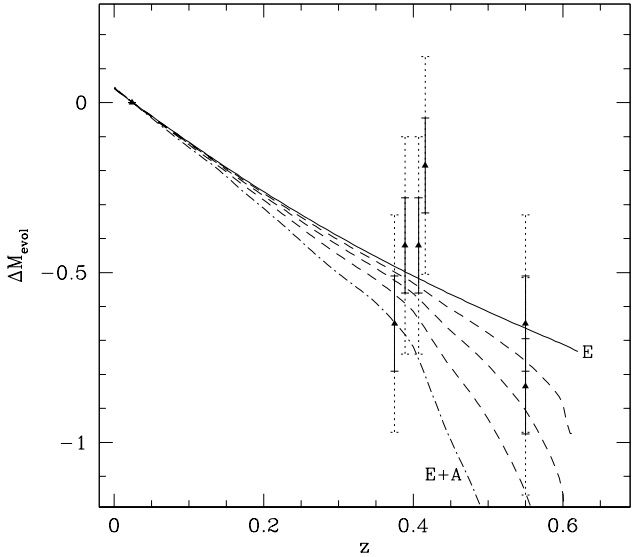,width=8.8cm}
\caption[]{Evolutionary BC98 model representing an elliptical galaxy that
  has experienced a second burst at $z=0.63$ as specified in the text
  (dot-dashed line) compared to the single 1-Gyr burst model (solid
  line). Dashed lines are 1:1, 1:4 and 1:10 combinations between the
  two models. }
\label{add}
\end{figure}

As a final example, we consider a scenario where an elliptical galaxy
experiences a sudden addition of a small second stellar population on
top of the old main component. A possible realization would be the
accretion of a small gas-rich galaxy leading to a second short burst of
star formation.  We have already seen that the contamination of a
sample of early-type galaxies by a few \ea s would not change
dramatically the Kormendy relation itself. Our \ea\ templates comprised
models with a second burst lasting 0.25 Gyrs, amounting to an
additional 20\% of the mass of
the underlying old population. A galaxy with a less prominent second
burst could easily be hidden in our ET samples.  The presence of the
second burst could be revealed by our spectrophotometric identification
if it occurred less than 2 Gyrs before $z_{\mathrm{obs}}$. If it
happened earlier, the \ea s signatures would not have been detected by our
method.  From the numerous possible model realizations of such an event
we pick up an example which lies close to our detection limit of a
second burst: we choose a model galaxy that had an initial 1-Gyr burst
of star formation lasting from $z=4$ until $z=2.6$ and that gets an
additional 10\% of mass in a second 0.2-Gyr burst at $z=0.63$,
corresponding to 2 Gyrs before $z=0.4$ in our cosmology.
Fig.\,\ref{add} shows that the difference in the luminosity evolution
at $z=0.4$ between this particular model and a pure burst model
corresponds to about 1 $\sigma_c$ error. We also plot the model
evolution for samples having different mixtures of these kind of
galaxies and single burst passively evolving galaxies which reduces the
difference from the pure burst model even further.

\section{Summary and conclusions}

We have investigated samples of elliptical galaxies in four clusters at
redshifts around $z=0.4$ and one at $z=0.55$. The cluster member
galaxies were selected by spectral type from our ground-based
spectrophotometric observations that allowed the classification of the
galaxies as ellipticals, spirals, irregulars and \ea s by comparing the
low-resolution SEDs with template spectra. The structural parameters
were determined from HST images by a two-component fitting of the
surface brightness profiles. With this method we derived not only
accurate values of the total magnitude and the effective radius of a
galaxy down to $B_{\mathrm{tot}}=23$\,\magn, but could also detect a
disk component if present down to the resolution limit and
derive disk-to-bulge values.

We constructed the rest-frame $B$-band Kormendy relations
($\mathrm{\langle\-\mu_e\rangle_{cor}}-\log(R_e)$) for the various
samples and find no significant change of the slope with
redshift. Because all the samples span a similar range in $\log(R_e)$,
this indicates that on average the stellar populations of smaller
ellipticals do not evolve in a dramatically different manner than
larger ones at $z<0.6$ implying a high redshift of formation for the
majority of the stars in early-type cluster galaxies irrespective of
the galaxies' size. The residuals of the Kormendy relations have a
rather high dispersion ($\sigma_{\mathrm{stat}}=0.14$\,\magn), which
is mainly due to having neglected the third parameter (the velocity
dispersion of the Fundamental Plane) and are little augmented by
measurement errors. The systematic errors arising from the calibration
of the HST magnitudes to rest-frame $B$ magnitudes (zeropoint and
color transformation, K-correction, and reddening) amount to the same
value.

\cbstart We have shown that the actual values of the derived
luminosity evolution depends on a number of different assumptions
starting with the choice of the local comparison sample. A further
assumption to be made is the appropriate 
magnitude cut-off for the local sample,
which must be restricted to the magnitude distribution of the
respective distant cluster in order to have an unbiased
comparison. For our Coma SBD sample we found that the variation of
this cut-off by half a magnitude results in differences of about
$0.05\,\magn$ in the estimated luminosity evolution of a distant cluster
sample. As mentioned in the Introduction, most authors using the
Kormendy relation take a fixed slope $b$ when fitting the data of
different clusters. We find that the derived luminosity evolution
remains the same within about $0.04\,\magn$ when the value of the
fixed slope is varied within the range found for local galaxy samples.\cbend{}

\cbstart Compared to our Coma SBD early-type galaxies we find for our initial
set of parameters an average brightening of $\Delta B=-0.42$\,\magn\
at $z=0.4$ and $\Delta\-B=-0.73$\,\magn\ at $z=0.55$. The scatter
between the four clusters at $z=0.4$ ($\sigma=0.15$\,\magn) is 
close to the statistical error for the
individual data point. Given the dependence of the derived brightening
on the various assumptions it is not surprising that other studies
find different values for the same clusters, especially when taking
into account that already the calibration of the HST images are performed
using other values for the zeropoint, K-correction, and
extinction. For example, Schade et al. \cite*{SCYLE96} give
$\Delta\-B=-0.22\pm0.19$\,\magn\ for a sample of 6 galaxies in the $I$
frame of \adsn\ and $\Delta\-B=-0.57\pm0.13$\,\magn\ for 28
ellipticals in the $I$ frame of \nnes\ assuming a fixed slope
$b=3.33$. Barger et al. \cite*{BASECDOPS98} get $\Delta
B=-0.45\pm0.09$\,\magn\ (judging from their Figure\,5b) for three
clusters at $z=0.4$ and $\Delta\-B=-0.62\pm0.1$\,\magn\ for three
clusters at $z=0.55$ assuming a fixed slope $b=3.0$ and no colour
dependence of the HST zeropoints.  \cbend{}

Contrary to previous studies, we could reliably detect the
post-starburst nature of a galaxy from our spectrophotometry and
exclude these galaxies from the samples of early-type galaxies. If we
contaminate these samples by the few \ea\ galaxies found in the core of
the clusters (about 10\% of the whole sample) we still do not find any
excess brightening of the Kormendy relations but only variations within
the 1\,$\sigma_c$ error. This is in accordance with model expectations
by Barger et al. \cite*{BAECSS96} who find only a small average
brightening of the spheroidal galaxy population of clusters at $z=0.3$
by the inclusion of \ea\ galaxies.

There is also no systematic trend towards stronger or weaker evolution
when we subdivide the samples into early-type galaxies having larger or
lower disk-to-bulge values than $d/b=0.2$. This indicates that most of
the (spectroscopically selected) galaxies with prominent disks in the
cores of distant clusters have disk populations consisting still mainly
of old stars. There is no significant contribution to the light by
young stars as would be expected if those galaxies were the recent
remnants of spiral galaxies that lost their gas due to some cluster
influence.

The presented luminosity evolution of early-type galaxies since intermediate
redshifts as derived here from the Kormendy relations is compatible with
passive evolution models. But given the relatively large total error (not
always considered in the past), the models are not constrained very much.
All burst models with formation redshift $z_f>2$ can reasonably fit the data.
Models with an exponentially decreasing star formation rate are also
adequate, as
long as the e-folding timescale $\tau$ is less than 2. We have shown that
galaxies with a younger formation epoch or a weak second burst of star
formation could be easily hidden in the Kormendy relation.

All in all, it is evident that the comparison of the Kormendy relations
at various redshifts does not constrain the luminosity evolution of
cluster ellipticals strongly enough to be able to decide whether pure passive
evolutionary models or models with exponentially decaying star
formation (which would fit better to hierarchical galaxy formation) 
can better match the data at
redshifts up to $z=0.6$. The significance of this method would be
increased if more observations of clusters at many other (and higher)
redshifts were combined although the internal scatter per cluster would
not be decreased. For few clusters,
 the investigation of the evolution of the tight
Fundamental Plane in connection with the Mg-$\sigma$ relationship
gives more accurate and constraining results.  


\begin{acknowledgements}
This research was partially supported by the Sonderforschungsbereich
375 and DARA grant 50\,OR\,9608\,5. \cbstart Some image reduction was
done using the MIDAS and/or IRAF/STSDAS packages. 
IRAF is distributed by the National Optical
Astronomy Observatories, which are operated by the Association of
Universities for Research in Astronomy, Inc., under cooperative
agreement with the National Science Foundation. \cbend{}
\end{acknowledgements}




\appendix

\section{Photometric parameters of the distant galaxy samples}
\label{photpar}

In this appendix we present the photometric parameters for all the
investigated galaxies, not only the early-type galaxies, which are
members of the distant clusters of this study. The parameters were
derived by the method described and extensively tested by Saglia et
al. \cite*{SBBBCDMW97}, which applies a PSF broadened 2-component
($r^{1/4}$ and exponential) fitting procedure. There is a separate
table for each cluster sample of Table\,\ref{gals}.

\textit{Explanation of the table columns:} \\
Col.\,1 (\textsf{galaxy}):
Identification number of the galaxy. For clusters \nndn\ and \nnes,
they correspond to the ID number of Belloni and R\"oser
\cite*{BR96}. For \adsn, they correspond to the ID number of
Soucail et al. \cite*{SMFC88}. For clusters \evvs\ and \ndnd\ (and \adsn), 
galaxy identifications of Smail et al. \cite*[S97]{SDCEOBS97} are
given, too. \\ 
Col.\,2 (\textsf{type}): The
numbers correspond to the SED model which best fits the
spectrophotometric data of the galaxy as described in Belloni et al.
\cite*{BBTR95}: $1=$ early-type (E, S0, or Sa), $2=$ spiral (Sbc), 
$3=$ spiral (Scd), $4=$ irregular (Im), 
$5-10=$ post-starburst (\ea) model. \\ 
Col.\,3 ($r_e$): Global effective radius in arcsec. \\
Col.\,4 ($\log R_e$): Global effective radius in kpc for
$H_0=60\,\mathrm{km\,s\me\,Mpc\me}, q_0=0.1$. \\
Col.\,5 ($M_{\mathrm tot}$): Total magnitude transformed to the
respective Johnson--Cousin magnitude and corrected for galactic
extinction according to SFD 
($M_{\mathrm tot}=M_{\mathrm inst}+{\mathrm ZP}+A$). See
Table\,\ref{calib} for the values of ZP and $A$(SFD) for the different
cluster samples. \\
Col.\,6 ($B_{\mathrm rest}$): Total magnitude transformed to
restframe Johnson $B$ ($B_{\mathrm rest}=M_{\mathrm tot}-K$). See
Table\,\ref{calib} for the values of $K$ for the different
cluster samples. \\
Col.\,7 ($\langle\mu_e\rangle_{\mathrm cor}$): Effective mean surface
brightness within $r_e$ in $B$ and corrected for the cosmological
surface brightness dimming (see equation\,\ref{sbcor}). \\
Col.\,8 ($d/b$): Disk-to-bulge ratio ($F_{\mathrm disk}/F_{\mathrm bulge}$). \\
Col.\,9 ($r_{e,b}$): Effective radius of the bulge component in
arcsec. \\
Col.\,10 ($h$): Disk scale length in arcsec. \\

\begin{table*}
\caption[]{Photometric parameters for \textit{a370i}}
\begin{flushleft}
\begin{tabular}{rrlllllllrr}
\noalign{\smallskip}
\hline
\noalign{\smallskip}
galaxy & S97 & type & $r_e$ & $\log R_e$ & $M_{\mathrm tot}$ & $B_{\mathrm rest}$ & $\langle\mu_e\rangle_{\mathrm cor}$ & $d/b$ & $r_{e,b}$ & $h$ \\
\noalign{\smallskip}
\hline
\noalign{\smallskip}
 38 &514&  1 &  0.51 & 0.45 & 18.87 & 20.75 & 19.90 & 0.00 &  0.51 & 0.00 \\ 
 46 &373&  2 &  0.85 & 0.68 & 18.74 & 20.62 & 20.87 & 2.44 &  0.34 & 0.62 \\ 
 47 &487&  1 &  1.36 & 0.88 & 18.23 & 20.11 & 21.38 & 0.22 &  1.00 & 2.46 \\ 
 49 &377&  1 &  1.40 & 0.89 & 18.22 & 20.10 & 21.44 & 0.22 &  1.06 & 2.02 \\ 
 59 &480&  2 &  1.35 & 0.88 & 17.88 & 19.76 & 21.03 & 1.19 &  0.89 & 0.98 \\ 
 72 &509&  2 &  0.69 & 0.58 & 19.57 & 21.45 & 21.24 & 1.38 &  0.53 & 0.45 \\ 
 -  &232&  1 &  0.64 & 0.56 & 18.56 & 20.43 & 20.09 & 0.52 &  0.50 & 0.50 \\ 
 -  &230&  1 &  0.55 & 0.49 & 18.41 & 20.28 & 19.60 & 0.00 &  0.55 & 0.00 \\ 
 -  &237&  1 &  1.21 & 0.83 & 18.70 & 20.58 & 21.61 & 0.00 &  1.21 & 0.00 \\ 
 -  &182&  1 &  1.05 & 0.77 & 18.07 & 19.95 & 20.68 & 0.00 &  1.05 & 0.00 \\ 
 -  &231&  1 &  0.48 & 0.43 & 19.14 & 21.02 & 20.05 & 0.00 &  0.48 & 0.00 \\ 
 -  &289&  1 &  0.57 & 0.51 & 19.41 & 21.29 & 20.70 & 0.00 &  0.57 & 0.00 \\ 
\noalign{\smallskip}
\hline
\end{tabular}
\end{flushleft}
\end{table*}

\begin{table*}
\caption[]{Photometric parameters for \textit{a370r}}
\begin{flushleft}
\begin{tabular}{rlllllllrr}
\noalign{\smallskip}
\hline
\noalign{\smallskip}
galaxy & type & $r_e$ & $\log R_e$ & $M_{\mathrm tot}$ & $B_{\mathrm rest}$ & $\langle\mu_e\rangle_{\mathrm cor}$ & $d/b$ & $r_{e,b}$ & $h$ \\
\noalign{\smallskip}
\hline
\noalign{\smallskip}
  7 &  1 &  0.42 & 0.37 & 20.21 & 21.23 & 19.95 & 0.00 &  0.42 & 0.00 \\ 
  9 &  6 &  4.95 & 1.44 & 18.82 & 19.84 & 23.92 & 1.42 &  3.09 & 3.54 \\ 
 13 &  1 &  0.83 & 0.67 & 18.88 & 19.90 & 20.11 & 0.00 &  0.83 & 0.00 \\ 
 16 &  1 &  0.88 & 0.69 & 19.04 & 20.06 & 20.40 & 0.51 &  2.18 & 0.19 \\ 
 17 &  1 &  1.21 & 0.83 & 18.69 & 19.71 & 20.73 & 0.49 &  0.68 & 1.63 \\ 
 18 &  1 &  0.88 & 0.69 & 19.04 & 20.06 & 20.40 & 0.35 &  0.56 & 1.33 \\ 
 20 &  1 &  7.63 & 1.63 & 16.59 & 17.61 & 22.63 & 2.81 &  1.99 & 5.80 \\ 
 23 &  1 &  1.39 & 0.89 & 18.48 & 19.50 & 20.83 & 0.37 &  0.89 & 1.86 \\ 
 27 &  1 &  0.52 & 0.47 & 19.51 & 20.53 & 19.73 & 0.00 &  0.52 & 0.00 \\ 
 28 &  1 &  1.53 & 0.93 & 18.45 & 19.47 & 21.00 & 0.16 &  1.89 & 0.44 \\ 
 31 &  1 &  0.78 & 0.64 & 19.28 & 20.30 & 20.37 & 0.66 &  0.32 & 1.18 \\ 
 32 &  1 &  1.89 & 1.02 & 18.27 & 19.29 & 21.29 & 0.55 &  0.96 & 2.67 \\ 
 35 &  1 &  9.13 & 1.71 & 16.73 & 17.75 & 23.16 & 0.07 & 10.36 & 0.76 \\ 
 36 &  1 &  0.46 & 0.41 & 19.83 & 20.85 & 19.77 & 0.00 &  0.46 & 0.00 \\ 
 56 &  1 &  0.44 & 0.39 & 19.40 & 20.42 & 19.23 & 0.15 &  0.35 & 0.74 \\ 
 64 &  1 &  0.47 & 0.41 & 20.07 & 21.09 & 20.04 & 0.00 &  0.47 & 0.00 \\ 
 70 &  6 &  0.46 & 0.41 & 20.15 & 21.17 & 20.10 & $\infty$ &   0.00 &  0.28 \\ 
 76 &  1 &  0.75 & 0.63 & 19.67 & 20.69 & 20.69 & 0.00 &  0.75 & 0.00 \\ 
 83 &  1 &  0.38 & 0.33 & 20.13 & 21.15 & 19.67 & 0.36 &  0.47 & 0.17 \\ 
\noalign{\smallskip}
\hline
\end{tabular}
\end{flushleft}
\end{table*}

\begin{table*}
\caption[]{Photometric parameters for \textit{a370v}}
\begin{flushleft}
\begin{tabular}{rrlllllllrr}
\noalign{\smallskip}
\hline
\noalign{\smallskip}
galaxy & S97 & type & $r_e$ & $\log R_e$ & $M_{\mathrm tot}$ & $B_{\mathrm rest}$ & $\langle\mu_e\rangle_{\mathrm cor}$ & $d/b$ & $r_{e,b}$ & $h$ \\
\noalign{\smallskip}
\hline
\noalign{\smallskip}
 38 &514&  1 &  0.56 & 0.49 & 20.68 & 20.43 & 19.78 & 0.00 &  0.56 & 0.00 \\ 
 46 &373&  2 &  0.85 & 0.68 & 20.43 & 20.18 & 20.43 & 3.25 &  0.28 & 0.60 \\ 
 47 &487&  1 &  1.49 & 0.92 & 19.97 & 19.72 & 21.20 & 0.08 &  1.72 & 0.11 \\ 
 49 &377&  1 &  1.13 & 0.80 & 20.33 & 20.08 & 20.95 & 0.00 &  1.13 & 0.00 \\ 
 59 &480&  2 &  1.31 & 0.86 & 19.71 & 19.46 & 20.65 & 3.06 &  0.50 & 0.92 \\ 
 72 &509&  2 &  0.61 & 0.53 & 20.73 & 20.48 & 20.01 & 2.75 &  0.24 & 0.43 \\ 
 -  &232&  1 &  0.63 & 0.54 & 20.57 & 20.32 & 19.92 & 0.96 &  0.38 & 0.50 \\ 
 -  &230&  1 &  0.57 & 0.50 & 19.90 & 19.65 & 19.04 & 0.00 &  0.57 & 0.00 \\ 
 -  &237&  1 &  1.19 & 0.82 & 20.57 & 20.32 & 21.32 & 0.00 &  1.19 & 0.00 \\ 
 -  &182&  1 &  0.97 & 0.73 & 19.97 & 19.72 & 20.26 & 0.00 &  0.97 & 0.00 \\ 
 -  &231&  1 &  0.50 & 0.44 & 20.98 & 20.73 & 19.83 & 0.00 &  0.50 & 0.00 \\ 
 -  &289&  1 &  0.76 & 0.63 & 21.12 & 20.87 & 20.90 & 0.17 &  1.04 & 0.10 \\ 
\noalign{\smallskip}
\hline
\end{tabular}
\end{flushleft}
\end{table*}

\begin{table*}
\caption[]{Photometric parameters for \textit{cl0016i}}
\begin{flushleft}
\begin{tabular}{rlllllllrr}
\noalign{\smallskip}
\hline
\noalign{\smallskip}
galaxy & type & $r_e$ & $\log R_e$ & $M_{\mathrm tot}$ & $B_{\mathrm rest}$ & $\langle\mu_e\rangle_{\mathrm cor}$ & $d/b$ & $r_{e,b}$ & $h$ \\
\noalign{\smallskip}
\hline
\noalign{\smallskip}
 40 &  1 &  0.38 & 0.41 & 20.14 & 21.76 & 19.73 & 0.18 &  0.49 & 0.10 \\ 
 43 &  4 &  0.26 & 0.25 & 21.07 & 22.69 & 19.85 & $\infty$ &   0.00 & 0.6 \\ 
 48 &  1 &  0.91 & 0.79 & 20.00 & 21.62 & 21.50 & 0.39 &  0.91 & 0.54 \\ 
 51 &  1 &  0.29 & 0.29 & 20.43 & 22.05 & 19.42 & 0.00 &  0.29 & 0.00 \\ 
 56 &  1 &  0.34 & 0.37 & 20.37 & 21.99 & 19.76 & 0.00 &  0.34 & 0.00 \\ 
 70 &  1 &  0.33 & 0.36 & 20.90 & 22.52 & 20.24 & 0.60 &  0.33 & 0.20 \\ 
 73 &  7 &  0.41 & 0.44 & 20.10 & 21.72 & 19.85 & 1.11 &  0.10 & 0.51 \\ 
 95 &  1 &  0.41 & 0.44 & 20.55 & 22.17 & 20.30 & 0.00 &  0.41 & 0.00 \\ 
 97 &  1 &  0.64 & 0.64 & 19.99 & 21.61 & 20.74 & 0.55 &  0.71 & 0.35 \\ 
122 &  1 &  0.48 & 0.52 & 19.19 & 20.81 & 19.31 & 0.00 &  0.48 & 0.00 \\ 
126 &  1 &  0.34 & 0.36 & 19.84 & 21.46 & 19.19 & 0.00 &  0.34 & 0.00 \\ 
133 &  1 &  0.52 & 0.55 & 20.07 & 21.69 & 20.35 & 0.54 &  0.24 & 0.89 \\ 
139 &  1 &  3.52 & 1.38 & 18.18 & 19.80 & 22.62 & 0.12 &  4.29 & 0.64 \\ 
141 &  3 &  0.77 & 0.72 & 19.42 & 21.04 & 20.58 & 0.32 &  1.29 & 0.19 \\ 
150 &  1 &  2.63 & 1.26 & 17.99 & 19.61 & 21.80 & 0.37 &  2.19 & 2.07 \\ 
152 &  1 &  2.64 & 1.26 & 17.85 & 19.47 & 21.67 & 0.29 &  3.55 & 0.93 \\ 
156 &  1 &  0.52 & 0.55 & 19.49 & 21.11 & 19.78 & 0.33 &  0.36 & 0.62 \\ 
160 &  1 &  0.64 & 0.64 & 19.61 & 21.23 & 20.34 & 0.00 &  0.64 & 0.00 \\ 
164 &  1 &  0.44 & 0.47 & 19.37 & 20.99 & 19.28 & 0.00 &  0.44 & 0.00 \\ 
175 &  1 &  0.33 & 0.36 & 20.68 & 22.30 & 20.01 & 0.00 &  0.33 & 0.00 \\ 
176 &  1 &  0.39 & 0.43 & 20.73 & 22.35 & 20.41 & 0.00 &  0.39 & 0.00 \\ 
179 &  1 &  0.50 & 0.53 & 19.59 & 21.21 & 19.78 & 0.00 &  0.50 & 0.00 \\ 
180 &  1 &  0.38 & 0.41 & 20.16 & 21.78 & 19.75 & 0.00 &  0.38 & 0.00 \\ 
181 &  1 &  0.45 & 0.48 & 19.56 & 21.18 & 19.52 & 0.25 &  0.33 & 0.59 \\ 
185 &  1 &  1.15 & 0.90 & 19.03 & 20.65 & 21.04 & 0.87 &  0.50 & 1.23 \\ 
187 &  1 &  0.57 & 0.59 & 19.77 & 21.39 & 20.25 & 0.00 &  0.57 & 0.00 \\ 
188 &  1 &  0.27 & 0.27 & 20.39 & 22.01 & 19.28 & 0.00 &  0.27 & 0.00 \\ 
191 &  1 &  0.44 & 0.48 & 20.47 & 22.09 & 20.41 & 0.00 &  0.44 & 0.00 \\ 
193 &  1 &  0.48 & 0.51 & 19.82 & 21.44 & 19.93 & 0.00 &  0.48 & 0.00 \\ 
207 &  7 &  0.30 & 0.31 & 20.20 & 21.82 & 19.28 & 0.44 &  0.14 & 0.88 \\ 
215 &  1 &  0.29 & 0.29 & 20.71 & 22.33 & 19.72 & 0.00 &  0.29 & 0.00 \\ 
222 &  1 &  0.48 & 0.52 & 20.46 & 22.08 & 20.57 & 0.00 &  0.48 & 0.00 \\ 
234 &  7 &  0.39 & 0.43 & 20.44 & 22.06 & 20.13 & 0.00 &  0.39 & 0.00 \\ 
\noalign{\smallskip}
\hline
\end{tabular}
\end{flushleft}
\end{table*}

\begin{table*}
\caption[]{Photometric parameters for \textit{cl0016v}}
\begin{flushleft}
\begin{tabular}{rlllllllrr}
\noalign{\smallskip}
\hline
\noalign{\smallskip}
galaxy & type & $r_e$ & $\log R_e$ & $M_{\mathrm tot}$ & $B_{\mathrm rest}$ & $\langle\mu_e\rangle_{\mathrm cor}$ & $d/b$ & $r_{e,b}$ & $h$ \\
\noalign{\smallskip}
\hline
\noalign{\smallskip}
 40 &  1 &  0.41 & 0.45 & 22.47 & 21.54 & 19.71 & 0.00 &  0.41 & 0.00 \\ 
 48 &  1 &  1.35 & 0.97 & 21.86 & 20.93 & 21.67 & 1.72 &  0.62 & 1.03 \\ 
 51 &  1 &  0.38 & 0.42 & 22.35 & 21.42 & 19.42 & 0.00 &  0.38 & 0.00 \\ 
 56 &  1 &  0.33 & 0.36 & 22.69 & 21.76 & 19.47 & 0.00 &  0.33 & 0.00 \\ 
 68 &  1 &  0.42 & 0.45 & 21.24 & 20.31 & 18.50 & 0.00 &  0.42 & 0.00 \\ 
 70 &  1 &  0.84 & 0.76 & 22.54 & 21.61 & 21.32 & 0.00 &  0.84 & 0.00 \\ 
 73 &  7 &  0.47 & 0.50 & 22.14 & 21.21 & 19.64 & 0.98 &  0.16 & 0.54 \\ 
 84 &  7 &  0.61 & 0.62 & 22.20 & 21.27 & 20.30 & 2.28 &  0.18 & 0.48 \\ 
 86 &  7 &  0.32 & 0.34 & 23.11 & 22.18 & 19.81 & 0.55 &  0.65 & 0.10 \\ 
 87 &  1 &  0.28 & 0.28 & 22.88 & 21.95 & 19.28 & 0.00 &  0.28 & 0.00 \\ 
 92 &  7 &  0.45 & 0.48 & 22.44 & 21.51 & 19.84 & 0.63 &  0.90 & 0.15 \\ 
 95 &  1 &  0.49 & 0.52 & 22.72 & 21.79 & 20.31 & 0.00 &  0.49 & 0.00 \\ 
 97 &  1 &  0.69 & 0.68 & 22.29 & 21.36 & 20.66 & 0.47 &  0.92 & 0.30 \\ 
109 &  7 &  0.65 & 0.65 & 21.77 & 20.84 & 19.98 & 0.44 &  0.55 & 0.48 \\ 
112 &  7 &  1.19 & 0.91 & 20.78 & 19.85 & 20.31 & $\infty$ &   0.00 & 0.71\\ 
126 &  1 &  0.32 & 0.35 & 22.25 & 21.32 & 18.96 & 0.00 &  0.32 & 0.00 \\ 
139 &  1 &  3.64 & 1.40 & 20.58 & 19.65 & 22.55 & 0.11 &  4.38 & 0.60 \\ 
144 &  1 &  0.34 & 0.37 & 23.09 & 22.16 & 19.94 & 0.00 &  0.34 & 0.00 \\ 
146 &  1 &  0.27 & 0.27 & 22.97 & 22.04 & 19.29 & 0.95 &  0.71 & 0.10 \\ 
150 &  1 &  1.70 & 1.06 & 20.28 & 19.35 & 20.59 & 3.85 &  0.18 & 1.28 \\ 
152 &  1 &  1.97 & 1.13 & 20.64 & 19.71 & 21.27 & 0.34 &  2.87 & 0.66 \\ 
156 &  1 &  0.52 & 0.55 & 21.85 & 20.92 & 19.60 & 0.00 &  0.52 & 0.00 \\ 
160 &  1 &  0.71 & 0.69 & 21.90 & 20.97 & 20.33 & 0.00 &  0.71 & 0.00 \\ 
162 &  1 &  0.28 & 0.28 & 22.98 & 22.05 & 19.36 & 0.00 &  0.28 & 0.00 \\ 
164 &  1 &  0.45 & 0.49 & 21.68 & 20.75 & 19.10 & 0.00 &  0.45 & 0.00 \\ 
173 &  3 &  0.53 & 0.56 & 22.05 & 21.12 & 19.84 & 1.15 &  0.16 & 0.59 \\ 
175 &  1 &  0.42 & 0.46 & 22.78 & 21.85 & 20.05 & 0.00 &  0.42 & 0.00 \\ 
179 &  1 &  0.43 & 0.47 & 22.01 & 21.08 & 19.33 & 0.00 &  0.43 & 0.00 \\ 
180 &  1 &  0.40 & 0.43 & 22.50 & 21.57 & 19.64 & 0.00 &  0.40 & 0.00 \\ 
181 &  1 &  0.75 & 0.71 & 21.66 & 20.73 & 20.18 & 0.81 &  0.28 & 0.97 \\ 
184 &  1 &  0.68 & 0.67 & 22.11 & 21.18 & 20.44 & 2.16 &  0.21 & 0.54 \\ 
187 &  1 &  0.67 & 0.66 & 21.99 & 21.06 & 20.27 & 0.00 &  0.67 & 0.00 \\ 
188 &  1 &  0.27 & 0.27 & 22.68 & 21.75 & 19.00 & 0.00 &  0.27 & 0.00 \\ 
193 &  1 &  0.48 & 0.51 & 22.18 & 21.25 & 19.73 & 0.00 &  0.48 & 0.00 \\ 
206 &  1 &  0.38 & 0.42 & 22.82 & 21.89 & 19.89 & 0.00 &  0.38 & 0.00 \\ 
207 &  7 &  0.27 & 0.27 & 22.34 & 21.41 & 18.66 & 0.00 &  0.27 & 0.00 \\ 
215 &  1 &  0.36 & 0.40 & 22.83 & 21.90 & 19.80 & 0.00 &  0.36 & 0.00 \\ 
222 &  1 &  0.67 & 0.66 & 22.56 & 21.63 & 20.85 & 0.00 &  0.67 & 0.00 \\ 
274 &  3 &  0.36 & 0.40 & 21.56 & 20.63 & 18.52 & 0.70 &  0.80 & 0.12 \\ 
\noalign{\smallskip}						  
\hline
\end{tabular}
\end{flushleft}
\end{table*}

\begin{table*}
\caption[]{Photometric parameters for \textit{cl0303r}}
\begin{flushleft}
\begin{tabular}{rrlllllllrr}
\noalign{\smallskip}
\hline
\noalign{\smallskip}
galaxy & S97 & type & $r_e$ & $\log R_e$ & $M_{\mathrm tot}$ & $B_{\mathrm rest}$ & $\langle\mu_e\rangle_{\mathrm cor}$ & $d/b$ & $r_{e,b}$ & $h$ \\
\noalign{\smallskip}
\hline
\noalign{\smallskip}
145 &162&  1 &  0.43 & 0.40 & 20.57 & 21.50 & 20.13 & 0.00 &  0.43 & 0.00 \\ 
151 &241&  1 &  0.53 & 0.49 & 20.81 & 21.74 & 20.83 & 0.60 &  0.51 & 0.32 \\ 
153 &256&  1 &  0.62 & 0.57 & 19.25 & 20.18 & 19.63 & 0.00 &  0.62 & 0.00 \\ 
165 &292&  1 &  0.96 & 0.75 & 19.75 & 20.68 & 21.06 & 0.12 &  1.20 & 0.10 \\ 
172 &374&  1 &  2.27 & 1.13 & 18.34 & 19.27 & 21.54 & 0.04 &  2.43 & 0.40 \\ 
176 &337&  8 &  0.45 & 0.43 & 20.88 & 21.81 & 20.58 & 1.22 &  0.81 & 0.21 \\ 
190 &439 & 1 &  0.56 & 0.52 & 20.14 & 21.07 & 20.29 & 0.00 &  0.56 & 0.00 \\ 
203 &431 & 1 &  0.57 & 0.53 & 21.12 & 22.05 & 21.30 &13.03 &  0.10 & 0.36 \\ 
214 &495 & 1 &  1.52 & 0.95 & 19.98 & 20.91 & 22.30 & 0.07 &  1.72 & 0.10 \\ 
222 &508 & 1 &  0.77 & 0.66 & 20.33 & 21.26 & 21.18 & 0.73 &  0.33 & 0.98 \\ 
224 &545 & 1 &  1.00 & 0.77 & 19.40 & 20.33 & 20.80 & 0.16 &  0.78 & 2.23 \\ 
245 &647 & 1 &  0.41 & 0.38 & 21.13 & 22.06 & 20.58 & 0.00 &  0.41 & 0.00 \\ 
247 &674 & 1 &  0.26 & 0.19 & 20.96 & 21.89 & 19.44 & 0.24 &  0.32 & 0.10 \\ 
264 &769 & 1 &  0.32 & 0.28 & 21.25 & 22.18 & 20.21 & 0.00 &  0.32 & 0.00 \\ 
268 &761 & 1 &  0.39 & 0.37 & 21.03 & 21.96 & 20.43 & 0.00 &  0.39 & 0.00 \\ 
269 &755 & 1 &  0.42 & 0.40 & 20.69 & 21.62 & 20.23 & 0.33 &  0.48 & 0.21 \\ 
270 &2020&  5 &  2.78 & 1.22 & 20.44 & 21.38 & 24.08 & 1.02 &  5.33 &1.21 \\ 
278 &794 & 5 &  0.65 & 0.59 & 20.59 & 21.52 & 21.08 & $\infty$ &   0.00 & 0.39 \\ 
283 &2033&  1 &  0.64 & 0.58 & 20.02 & 20.95 & 20.47 & 0.82 &  0.48 &0.46 \\ 
290 &835 & 8 &  0.28 & 0.22 & 21.49 & 22.42 & 20.13 & 0.00 &  0.28 & 0.00 \\ 
297 &848 & 1 &  0.34 & 0.30 & 21.26 & 22.19 & 20.32 & 0.00 &  0.34 & 0.00 \\ 
301 &880 & 1 &  0.73 & 0.64 & 20.69 & 21.62 & 21.43 & 0.44 &  0.76 & 0.42 \\ 
307 &879 & 1 &  0.66 & 0.59 & 20.12 & 21.05 & 20.62 & 0.77 &  0.37 & 0.61 \\ 
316 &909 & 1 &  0.44 & 0.41 & 20.33 & 21.26 & 19.94 & 0.00 &  0.44 & 0.00 \\ 
318 &933 & 1 &  0.41 & 0.39 & 20.55 & 21.48 & 20.05 & 0.44 &  0.46 & 0.22 \\ 
327 &946 & 1 &  0.41 & 0.39 & 21.16 & 22.09 & 20.65 & 0.00 &  0.41 & 0.00 \\ 
329 &966 & 9 &  0.60 & 0.55 & 20.97 & 21.90 & 21.27 & 0.00 &  0.60 & 0.00 \\ 
344 &996 & 1 &  0.83 & 0.69 & 20.13 & 21.06 & 21.14 & 0.00 &  0.83 & 0.00 \\ 
361 &1025& 10 &  0.50 & 0.47 & 21.26 & 22.19 & 21.18 & 0.00 &  0.50 &0.00 \\ 
368 &1037&  1 &  0.79 & 0.67 & 19.63 & 20.56 & 20.53 & 0.97 &  0.52 &0.60 \\ 
\noalign{\smallskip}						      
\hline
\end{tabular}
\end{flushleft}
\end{table*}

\begin{table*}
\caption[]{Photometric parameters for \textit{cl0939i}}
\begin{flushleft}
\begin{tabular}{rlllllllrr}
\noalign{\smallskip}
\hline
\noalign{\smallskip}
galaxy & type & $r_e$ & $\log R_e$ & $M_{\mathrm tot}$ & $B_{\mathrm rest}$ & $\langle\mu_e\rangle_{\mathrm cor}$ & $d/b$ & $r_{e,b}$ & $h$ \\
\noalign{\smallskip}
\hline
\noalign{\smallskip}
615 &  1 &  0.46 & 0.43 & 20.07 & 21.91 & 20.73 & 0.70 &  0.36 & 0.33 \\ 
632 &  1 &  0.52 & 0.49 & 20.47 & 22.31 & 21.42 & 0.00 &  0.52 & 0.00 \\ 
646 &  7 &  0.63 & 0.57 & 18.48 & 20.32 & 19.82 & 0.24 &  0.50 & 0.70 \\ 
650 &  1 &  0.31 & 0.25 & 20.27 & 22.11 & 20.05 & 0.00 &  0.31 & 0.00 \\ 
663 &  1 &  0.51 & 0.48 & 19.57 & 21.41 & 20.46 & 0.00 &  0.51 & 0.00 \\ 
686 &  1 &  0.41 & 0.38 & 19.98 & 21.82 & 20.39 & 0.25 &  0.32 & 0.46 \\ 
688 &  8 &  0.67 & 0.59 & 19.58 & 21.42 & 21.04 & 0.22 &  0.83 & 0.24 \\ 
746 &  1 &  0.69 & 0.61 & 18.28 & 20.12 & 19.83 & 0.15 &  0.91 & 0.10 \\ 
\noalign{\smallskip}
\hline
\end{tabular}
\end{flushleft}
\end{table*}

\begin{table*}
\caption[]{Photometric parameters for \textit{cl0939r}}
\begin{flushleft}
\begin{tabular}{rlllllllrr}
\noalign{\smallskip}
\hline
\noalign{\smallskip}
galaxy & type & $r_e$ & $\log R_e$ & $M_{\mathrm tot}$ & $B_{\mathrm rest}$ & $\langle\mu_e\rangle_{\mathrm cor}$ & $d/b$ & $r_{e,b}$ & $h$ \\
\noalign{\smallskip}
\hline
\noalign{\smallskip}
173 &  7 &  0.71 & 0.62 & 20.90 & 21.85 & 21.62 & 2.58 &  0.50 & 0.46 \\ 
176 &  1 &  1.02 & 0.78 & 19.00 & 19.95 & 20.51 & 0.73 &  0.46 & 1.22 \\ 
180 &  1 &  0.42 & 0.39 & 20.83 & 21.78 & 20.40 & 0.72 &  0.31 & 0.31 \\ 
183 &  1 &  0.27 & 0.19 & 21.14 & 22.09 & 19.73 & 0.80 &  0.19 & 0.20 \\ 
189 &  7 &  0.60 & 0.55 & 20.17 & 21.12 & 20.53 & 2.41 &  0.68 & 0.35 \\ 
205 &  8 &  0.27 & 0.20 & 21.30 & 22.25 & 19.93 & 0.00 &  0.27 & 0.00 \\ 
216 &  1 &  0.40 & 0.37 & 21.22 & 22.17 & 20.70 & 1.11 &  0.64 & 0.19 \\ 
218 &  1 &  0.42 & 0.40 & 20.90 & 21.85 & 20.50 & 0.82 &  0.31 & 0.31 \\ 
224 &  1 &  0.39 & 0.36 & 20.41 & 21.36 & 19.83 & 0.05 &  0.42 & 0.10 \\ 
234 &  1 &  0.51 & 0.47 & 19.75 & 20.70 & 19.74 & 0.82 &  0.21 & 0.60 \\ 
240 &  2 &  0.66 & 0.58 & 20.50 & 21.45 & 21.05 & 1.81 &  0.79 & 0.37 \\ 
244 &  1 &  0.45 & 0.42 & 21.65 & 22.60 & 21.39 & 0.72 &  0.49 & 0.25 \\ 
247 &  8 &  0.42 & 0.39 & 21.20 & 22.15 & 20.79 & 2.88 &  0.74 & 0.23 \\ 
251 &  1 &  0.45 & 0.42 & 20.85 & 21.80 & 20.57 & 0.00 &  0.45 & 0.00 \\ 
259 &  2 &  0.33 & 0.28 & 20.60 & 21.55 & 19.65 & $\infty$ &   0.00 & 0.20 \\ 
267 &  1 &  0.74 & 0.64 & 19.53 & 20.48 & 20.34 & 0.06 &  0.68 & 1.22 \\ 
272 &  1 &  0.74 & 0.64 & 19.45 & 20.40 & 20.25 & 0.49 &  0.37 & 1.25 \\ 
273 &  7 &  0.78 & 0.66 & 18.64 & 19.59 & 19.57 & 0.08 &  0.90 & 0.10 \\ 
274 &  7 &  0.31 & 0.26 & 22.47 & 23.42 & 21.38 & 4.89 &  0.27 & 0.19 \\ 
279 &  1 &  0.51 & 0.48 & 21.42 & 22.37 & 21.42 & 0.00 &  0.51 & 0.00 \\ 
282 &  7 &  0.40 & 0.37 & 21.83 & 22.78 & 21.30 & 0.19 &  0.50 & 0.13 \\ 
290 &  1 &  0.37 & 0.33 & 21.52 & 22.47 & 20.81 & 0.83 &  0.41 & 0.20 \\ 
292 &  1 &  0.29 & 0.23 & 21.31 & 22.26 & 20.07 & 1.02 &  0.27 & 0.18 \\ 
293 &  2 &  0.42 & 0.39 & 21.12 & 22.07 & 20.71 & 0.90 &  0.24 & 0.36 \\ 
294 &  1 &  0.65 & 0.58 & 19.52 & 20.47 & 20.04 & 0.00 &  0.65 & 0.00 \\ 
299 &  1 &  2.33 & 1.13 & 18.39 & 19.34 & 21.68 & 0.03 &  2.46 & 0.27 \\ 
304 &  1 &  2.06 & 1.08 & 18.38 & 19.33 & 21.40 & 0.07 &  2.35 & 0.16 \\ 
305 &  1 &  0.38 & 0.34 & 21.45 & 22.40 & 20.80 & 0.00 &  0.38 & 0.00 \\ 
345 &  3 &  0.79 & 0.67 & 21.11 & 22.06 & 22.06 & 0.32 &  1.05 & 0.30 \\ 
375 &  1 &  1.49 & 0.94 & 18.30 & 19.25 & 20.63 & 2.27 &  0.60 & 1.10 \\ 
378 &  1 &  0.61 & 0.56 & 19.96 & 20.91 & 20.37 & 0.29 &  0.63 & 0.35 \\ 
387 &  1 &  0.47 & 0.44 & 20.03 & 20.98 & 19.83 & 0.31 &  0.58 & 0.19 \\ 
388 &  1 &  0.34 & 0.29 & 22.01 & 22.96 & 21.10 & 0.65 &  0.67 & 0.12 \\ 
392 &  1 &  0.46 & 0.43 & 20.84 & 21.79 & 20.61 & 0.00 &  0.46 & 0.00 \\ 
414 &  2 &  0.48 & 0.45 & 21.71 & 22.66 & 21.59 & 0.00 &  0.48 & 0.00 \\ 
441 &  1 &  0.38 & 0.35 & 22.25 & 23.20 & 21.62 & 4.08 &  0.10 & 0.27 \\ 
480 &  1 &  0.32 & 0.27 & 20.17 & 21.12 & 19.14 & 0.65 &  0.14 & 0.45 \\ 
526 &  1 &  0.30 & 0.25 & 20.46 & 21.41 & 19.33 & 0.00 &  0.30 & 0.00 \\ 
572 &  7 &  0.45 & 0.42 & 20.68 & 21.63 & 20.40 & 1.91 &  0.25 & 0.31 \\ 
586 &  3 &  0.49 & 0.46 & 20.46 & 21.41 & 20.37 & 1.44 &  1.25 & 0.21 \\ 
589 &  7 &  0.32 & 0.27 & 21.09 & 22.04 & 20.07 & 0.58 &  0.50 & 0.13 \\ 
\noalign{\smallskip}
\hline
\end{tabular}
\end{flushleft}
\end{table*}

\begin{table*}
\caption[]{Photometric parameters for \textit{cl0939v}}
\begin{flushleft}
\begin{tabular}{rlllllllrr}
\noalign{\smallskip}
\hline
\noalign{\smallskip}
galaxy & type & $r_e$ & $\log R_e$ & $M_{\mathrm tot}$ & $B_{\mathrm rest}$ & $\langle\mu_e\rangle_{\mathrm cor}$ & $d/b$ & $r_{e,b}$ & $h$ \\
\noalign{\smallskip}
\hline
\noalign{\smallskip}
613 &  7 &  0.37 & 0.33 & 22.09 & 21.70 & 20.03 & 1.01 &  0.33 & 0.23 \\ 
615 &  1 &  0.53 & 0.49 & 21.91 & 21.52 & 20.65 & 0.34 &  0.50 & 0.35 \\ 
632 &  1 &  0.61 & 0.55 & 22.21 & 21.82 & 21.25 & 0.00 &  0.61 & 0.00 \\ 
646 &  7 &  0.77 & 0.65 & 20.12 & 19.73 & 19.67 & 0.42 &  0.48 & 0.93 \\ 
650 &  1 &  0.32 & 0.28 & 22.24 & 21.85 & 19.90 & 0.00 &  0.32 & 0.00 \\ 
663 &  1 &  0.63 & 0.57 & 21.32 & 20.93 & 20.45 & 0.24 &  0.96 & 0.12 \\ 
686 &  1 &  0.52 & 0.49 & 21.65 & 21.26 & 20.37 & 0.00 &  0.52 & 0.00 \\ 
688 &  7 &  1.60 & 0.97 & 20.93 & 20.54 & 22.07 & 0.26 &  2.61 & 0.20 \\ 
726 &  1 &  0.32 & 0.28 & 22.60 & 22.21 & 20.26 & 0.00 &  0.32 & 0.00 \\ 
731 &  1 &  0.57 & 0.52 & 22.43 & 22.04 & 21.33 & $\infty$ &   0.00 & 0.34 \\ 
746 &  1 &  0.65 & 0.58 & 20.16 & 19.77 & 19.34 & 0.00 &  0.65 & 0.00 \\ 
\noalign{\smallskip}
\hline
\end{tabular}
\end{flushleft}
\end{table*}

\begin{table*}
\caption[]{Photometric parameters for \textit{cl1447r}}
\begin{flushleft}
\begin{tabular}{rrlllllllrr}
\noalign{\smallskip}
\hline
\noalign{\smallskip}
galaxy & S97 & type & $r_e$ & $\log R_e$ & $M_{\mathrm tot}$ & $B_{\mathrm rest}$ & $\langle\mu_e\rangle_{\mathrm cor}$ & $d/b$ & $r_{e,b}$ & $h$ \\
\noalign{\smallskip}
\hline
\noalign{\smallskip}
 80 &488&  8 &  0.53 & 0.48 & 21.79 & 22.79 & 21.98 & 0.00 &  0.53 & 0.00 \\ 
 83 &216&  1 &  0.34 & 0.29 & 20.20 & 21.20 & 19.44 & 0.00 &  0.34 & 0.00 \\ 
 90 &164&  1 &  0.37 & 0.33 & 20.93 & 21.93 & 20.35 & 0.00 &  0.37 & 0.00 \\ 
102 &341&  1 &  0.38 & 0.33 & 20.58 & 21.58 & 20.03 & 0.19 &  0.49 & 0.10 \\ 
110 &452&  1 &  0.31 & 0.24 & 20.91 & 21.91 & 19.91 & 0.00 &  0.31 & 0.00 \\ 
112 &335&  1 &  1.17 & 0.83 & 19.50 & 20.50 & 21.41 & 0.33 &  0.89 & 1.17 \\ 
115 &282&  1 &  0.43 & 0.39 & 19.95 & 20.95 & 19.67 & 0.00 &  0.43 & 0.00 \\ 
116 &168&  1 &  0.33 & 0.27 & 20.80 & 21.80 & 19.93 & 0.00 &  0.33 & 0.00 \\ 
122 &539&  1 &  0.42 & 0.38 & 20.10 & 21.10 & 19.79 & 0.00 &  0.42 & 0.00 \\ 
125 &446&  1 &  0.68 & 0.59 & 19.39 & 20.39 & 20.10 & 0.32 &  0.46 & 0.93 \\ 
133 &351&  1 &  0.46 & 0.42 & 20.17 & 21.17 & 20.05 & 0.38 &  0.41 & 0.32 \\ 
134 &2010&  1 &  0.49 & 0.45 & 19.98 & 20.98 & 20.01 & 0.00 &  0.49& 0.00 \\ 
135 &56&  1 &  0.33 & 0.27 & 20.56 & 21.56 & 19.69 & 0.00 &  0.33  & 0.00 \\ 
154 &317&  1 &  0.35 & 0.30 & 20.66 & 21.66 & 19.95 & 0.30 &  0.52 & 0.10 \\ 
160 &2004&  1 &  1.13 & 0.81 & 19.03 & 20.03 & 20.87 & 0.20 &  1.44& 0.35 \\ 
168 &490&  1 &  1.09 & 0.79 & 19.84 & 20.84 & 21.58 & 0.10 &  1.29 & 0.15 \\ 
184 &301&  1 &  0.54 & 0.49 & 19.95 & 20.95 & 20.18 & 0.22 &  0.62 & 0.24 \\ 
185 &108&  1 &  0.94 & 0.73 & 20.63 & 21.63 & 22.07 & 0.34 &  1.73 & 0.17 \\ 
188 &2021&  1 &  1.19 & 0.83 & 20.21 & 21.21 & 22.16 & 2.65 &  0.24& 0.96 \\ 
200 &473&  1 &  2.04 & 1.07 & 18.44 & 19.44 & 21.56 & 0.36 &  1.30 & 2.97 \\ 
201 &413&  1 &  0.75 & 0.63 & 19.04 & 20.04 & 20.00 & 0.23 &  0.56 & 1.12 \\ 
214 &197&  1 &  0.69 & 0.59 & 20.90 & 21.90 & 21.65 & 0.00 &  0.69 & 0.00 \\ 
216 &237&  1 &  0.44 & 0.40 & 20.47 & 21.47 & 20.26 & 0.00 &  0.44 & 0.00 \\ 
217 &505&  1 &  0.28 & 0.21 & 20.57 & 21.57 & 19.38 & 0.00 &  0.28 & 0.00 \\ 
220 &549&  1 &  0.46 & 0.42 & 20.88 & 21.88 & 20.76 & 0.00 &  0.46 & 0.00 \\ 
231 &153&  1 &  0.44 & 0.40 & 21.21 & 22.21 & 20.98 & 0.00 &  0.44 & 0.00 \\ 
247 &812&  9 &  0.47 & 0.43 & 20.28 & 21.28 & 20.19 & 0.15 &  0.60 & 0.10 \\ 
257 &572&  1 &  0.43 & 0.39 & 20.80 & 21.80 & 20.54 & 0.00 &  0.43 & 0.00 \\ 
283 &755&  1 &  0.32 & 0.26 & 20.77 & 21.77 & 19.85 & 0.00 &  0.32 & 0.00 \\ 
288 &729&  1 &  0.68 & 0.59 & 19.93 & 20.93 & 20.66 & 1.66 &  0.33 & 0.51 \\ 
318 &671&  1 &  0.43 & 0.39 & 21.35 & 22.35 & 21.08 & 0.74 &  0.27 & 0.36 \\ 
323 &709&  1 &  0.80 & 0.66 & 19.57 & 20.57 & 20.65 & 0.00 &  0.80 & 0.00 \\ 
339 &516&  1 &  1.08 & 0.79 & 20.07 & 21.07 & 21.81 & 0.19 &  1.44 & 0.26 \\ 
\noalign{\smallskip}
\hline
\end{tabular}
\end{flushleft}
\end{table*}





\end{document}